\documentclass[10pt]{article}
\usepackage{upgreek}
\usepackage[utf8]{inputenc}
\usepackage[english]{babel}
\usepackage{graphicx,amsmath,amsfonts,amssymb,dcolumn,amsthm,slashed,bbm,xspace,pgffor,tikz,mathbbol,latexsym,cite,braket}

\usepackage{microtype,tikz}
\usepackage{lmodern}
\usepackage[normalem]{ulem}
\usepackage{xcolor}
\usepackage[colorlinks=true,citecolor=blue,urlcolor=blue,linkcolor=blue]{hyperref}
\usepackage[hang]{footmisc} 
\usepackage[affil-it]{authblk}
\usepackage[compat=1.1.0]{tikz-feynhand}
\usepackage{multicol}
\usepackage{microtype,diagbox}

\newcommand{\orcidicon}{%
	\begin{tikzpicture}
	\draw[lime, fill=lime] (0,0)
		circle [radius=0.16]
		node[white] {{\fontfamily{qag}\selectfont \tiny ID}};
	\draw[white, fill=white] (-0.0625,0.095)
		circle [radius=0.007];
	\end{tikzpicture}	\hspace{-2mm}
}

\newcommand\orcidroldao{{\href{https://orcid.org/0000-0003-3978-532X}{\orcidicon}}}
\newcommand\orcido{{\href{https://orcid.org/0000-0002-2685-3349}{\orcidicon}}}

\numberwithin{equation}{section}

\usepackage{float}
\begin{document}

\title{\textbf{Configurational entropy, thermodynamic instability and phase transitions in holographic QCD}}

\author[1,2]{Octavio C. Junqueira\protect\orcido\thanks{\href{mailto:octavioj@pos.if.ufrj.br}{octavioj@pos.if.ufrj.br}}}
\author[1]{Roldao da Rocha\protect\orcidroldao\thanks{\href{mailto:roldao.rocha@ufabc.edu.br}{roldao.rocha@ufabc.edu.br}}}
\affil[1]{Center of Mathematics, Federal University of ABC, 09210-580\\ Santo Andr\'e, S\~ao Paulo, Brazil}
\affil[2]{Center of Physics, Federal University of ABC, 09210-580\\ Santo Andr\'e, S\~ao Paulo, Brazil}

\date{}
\maketitle

\begin{abstract}

We investigate the configurational entropy (CE) of rotating
strongly interacting matter at finite density within
Andreev's soft-wall AdS/QCD model. Holographically
renormalized thermodynamic energies are employed to
construct radial energy distributions for the confined
and deconfined phases. For the holographic energy density prescribed here, the CE remains invariant throughout the hadronic phase, exhibits a discontinuity across first-order confinement/deconfinement transitions, and varies smoothly in thermodynamic ensembles displaying crossover-like behavior. Comparing the CE with the specific heat shows that configurational stability is not necessarily associated with local thermodynamic stability, as determined by the sign of the specific heat. We show that a pronounced maximum of the ensemble CE emerges near the critical region of the 
QCD phase diagram, without a corresponding anomaly in
the specific heat, pointing to an information-theoretic signature of the critical region separating first-order transitions from crossover behavior.

\end{abstract}

\section{Introduction}
\label{SecIntro}

The characterization of phase transitions in strongly interacting
matter involves more than establishing the thermodynamic conditions
under which confinement or deconfinement takes place. It also
requires understanding how the redistribution of energy among the
available degrees of freedom modifies the spectral redistribution of the radial energy density across the confinement/deconfinement transition. The thermodynamic properties of the quark-gluon plasma
(QGP), produced in ultrarelativistic heavy-ion collisions, have
been extensively investigated at RHIC and the LHC, with complementary
perspectives emerging from the experimental programs at FAIR and
NICA \cite{Critelli:2017oub,Shuryak:2008eq,Busza:2018rrf}.
Nevertheless, the mechanisms governing the rearrangement of the
degrees of freedom across the quantum chromodynamics (QCD) phase
diagram remain insufficiently characterized, particularly at finite
baryon density and in the presence of collective rotation \cite{Ferreira:2025iqe}.
The existence and location of the QCD critical point, as well as
the physical mechanisms underlying the change in the order of the
confinement/deconfinement transition, constitute central aspects
of this problem
\cite{Fukushima:2010bq,Toniato:2025gts,Bzdak:2019pkr}.
Although lattice QCD calculations establish the crossover nature
of the transition at vanishing baryon chemical potential
\cite{Aoki:2006we,Borsanyi:2010bp,HotQCD:2018pds,Cucchieri:2007rg}, the sign problem restricts the extension of
first-principles calculations to finite-density regimes
\cite{Muroya:2003qs}, motivating complementary nonperturbative
approaches \cite{Krein:2017usp}.

An aspect of the QCD phase structure that deserves particular
attention is the distinction between thermodynamic and
configurational properties of strongly interacting matter.
Thermodynamic potentials and their derivatives characterize
the macroscopic response of a system to changes in its control
parameters, without explicitly resolving the relative
contributions of the modes composing its energy distribution \cite{MartinContreras:2020cyg}.
The latter carry information concerning the organization of
physical configurations that is not necessarily accessible
through conventional thermodynamic observables. This distinction
becomes particularly relevant when comparing the confined
hadronic phase, governed by color-singlet excitations, with
the deconfined phase, in which quark and gluon degrees of
freedom participate in the collective dynamics. Whether such
a rearrangement admits an independent informational
characterization, and whether this characterization can
distinguish first-order transitions from crossovers, are
questions naturally addressed within the framework of
configurational entropy (CE).

The concept of CE originates from Shannon's formulation of
information theory \cite{shannon} and its subsequent extension
to continuous physical configurations by Gleiser and Stamatopoulos
\cite{Gleiser:2011di,Gleiser:2012tu,Gleiser:2013mga}.
Its introduction into AdS/QCD 
\cite{Bernardini:2016hvx} established an information-theoretic
approach to hadronic physics, relating the spectral organization
of holographic energy distributions to the spectral hierarchy and configurational stability of mesonic resonances.
Subsequent developments encompassed 
heavy quarkonia and pseudoscalar meson families 
\cite{Bernardini:2018uuy,MartinContreras:2021bis,Zhao:2023yry,Ferreira:2019inu,MartinContreras:2023oqs,Lee:2021rag,daRocha:2024sjn,Karapetyan:2023kzs}.
The scope of configurational information measures was further
extended to holographic light-front wave functions and
nonperturbative aspects of QCD dynamics
\cite{Braga:2021fey,Colangelo:2018mrt,Karapetyan:2018oye,Karapetyan:2023sfo,Ferreira:2020iry,Wang:2025cam}, as well as to
the characterization of exotic hadronic configurations, holographic isospectral models, and the
spectroscopy and stability of light nuclei
\cite{
MartinContreras:2023eft,MartinContreras:2022lxl,Kou:2023azd}.
Complementary developments in bottom-up holographic QCD
have addressed baryon spectroscopy in the context of information entropy 
\cite{Guo:2024nrf,MartinContreras:2025mpj,Almeida:2026ozb,Diles:2025xot}. More recent approaches also take into account quantum information entropies to approach aspects of QCD \cite{Mahapatra:2019uql,daRocha:2025zth,daRocha:2021xwq,Jain:2022hxl,Kharzeev:2026inq,Rabelo-Soares:2025ams}. 
Together, these analyses establish the information-theoretic measures of hadronic structure and stability, while
providing additional criteria for addressing the
phenomenological consistency of holographic constructions.
Complementary developments in nuclear physics have also established
connections between information entropy and phase transitions \cite{Ma:2018wtw}.
Within holographic QCD at finite temperature, configurational
information measures have been employed to investigate quarkonium
dissociation, color confinement, deconfinement transitions,
and the stability of rotating plasmas
\cite{Braga:2020opg,Braga:2023qee}.
By assigning relative weights to the Fourier modes of a localized
energy distribution, the CE quantifies the informational content
associated with its spectral organization. Unlike thermodynamic
entropy, which accounts for the statistical multiplicity of
microscopic configurations, the CE depends on the relative
distribution of the modes constituting a prescribed physical
configuration \cite{Mvondo-She:2023xel}. Consequently, configurations with identical
integrated energies may have distinct informational contents,
whereas variations in the total energy that preserve the relative
spectral weights leave the CE invariant \cite{Bazeia:2021stz,Lima:2022ipl}. These properties enable the CE to characterize changes in the spectral structure of holographic energy distributions across the QCD phase diagram, 
particularly when changes in their spectral organization are not
manifest in conventional thermodynamic observables.

The gauge/gravity correspondence
\cite{Maldacena:1997re,Gubser:1998bc,Witten:1998qj}
provides a suitable framework for implementing this
information-theoretic description in strongly-coupled gauge
theories. In holographic QCD, the dependence of bulk configurations on the holographic coordinate encodes the renormalization-group flow of the dual field theory \cite{Anabalon:2023lnk}, allowing nonperturbative phenomena to be
investigated through their gravitational counterparts
\cite{Rougemont:2017tlu,Branz:2010ub,Colangelo:2008us,
Brodsky:2014yha,Rougemont:2023gfz}.
The introduction of an infrared (IR) scale into the bulk geometry,
as implemented in the soft-wall AdS/QCD model
\cite{Karch:2006pv,Gursoy:2007cb,Mamedov:2021hkb}, is particularly appropriate
for this purpose. The dilaton background sets a characteristic scale controlling the localization of bulk configurations along the holographic direction, while the corresponding Fourier spectra
encode information concerning their distribution along the
holographic direction. The phenomenological success of
bottom-up holographic constructions in describing confinement,
hadronic spectra, and chiral symmetry breaking has motivated
several developments and extensions of this framework
\cite{Ballon-Bayona:2017bwk,Bartz:2018nzn,
Ballon-Bayona:2023zal,Aref'eva2018May,Gherghetta:2009ac,Zhao:2026bsc,
Erlich:2005qh,Sakai2005Apr,Li:2013oda}.
Different realizations of the IR sector have also
provided complementary descriptions of hadronic and plasma
properties
\cite{Shukla:2023pbp,DaRold:2005mxj,Jena:2024cqs,Zhu:2026pqm}.

From the informational standpoint, a particularly relevant
feature of holographic deconfinement is the distinct extent of the holographic coordinate in the bulk geometries dual to the hadronic and plasma phases. Whereas the
thermal AdS geometry extends throughout the holographic
direction, the exterior region of the AdS black hole (BH)
is bounded by an event horizon whose location depends on
the thermodynamic parameters of the dual gauge theory.
The Hawking--Page (HP) transition, which holographically
implements the confinement/deconfinement transition
\cite{Noronha:2009ud,Herzog:2006ra,BallonBayona:2007vp},
therefore corresponds to a change in the dominant gravitational saddle between configurations whose energy distributions are supported on distinct domains of the holographic coordinate. Such a geometric distinction has
direct consequences for the corresponding Fourier spectra
and, hence, for their CEs. The extension
of previous configurational analyses of holographic
deconfinement and plasma stability
\cite{Braga:2020opg,Braga:2023qee}
to finite-density rotating matter introduces additional
features, since both the horizon position and the
thermodynamic contributions to the energy distribution
depend on the chemical potential and angular velocity.
The holographic direction thus provides a common geometric
setting in which the thermodynamic and configurational
properties of the competing phases can be examined without
identifying their respective notions of entropy.

Rotation provides an additional macroscopic parameter
through which the informational properties of strongly
interacting matter can be investigated. In noncentral
heavy-ion collisions, the angular momentum transferred
to the medium generates fluid vorticity
\cite{STAR:2017ckg}, whose physical manifestations include
the polarization of emitted hadrons and anomalous transport
phenomena, particularly the chiral vortical effect
\cite{Jiang:2016woz,Kharzeev:2015znc}.
The interplay between vorticity, viscous dissipation, and
the expansion of the QGP further modifies the collective
dynamics of the medium \cite{Abboud:2023hos}.
Developments in relativistic dissipative hydrodynamics 
provide a complementary perspective on the macroscopic
response of rotating strongly interacting matter \cite{Bemfica:2020xym,Rocha:2022ind,Kovtun:2019hdm}.
The configurational description pursued here, however,
addresses a different aspect of the system, namely the
redistribution of the spectral weights of its holographic
energy density as rotation modifies its thermodynamic state.

The influence of rotation on the QCD phase structure has
been investigated through phenomenological and holographic
approaches, addressing the dependence of the deconfinement
temperature on angular velocity, the emergence of
inhomogeneous configurations, and modifications of the
chiral condensate
\cite{Chen:2017xrj,Wang:2018sur,
Fujimoto:2021xix,Zhao:2022uxc}.
Lattice QCD investigations have also revealed a nontrivial
interplay between rotation and confinement
\cite{Braguta:2023iyx}, whereas holographic studies have
explored its effects on the deconfinement temperature
and the informational properties of the QGP
\cite{Braga:2022yfe,Braga:2025eiz}.
These developments suggest that rotation may influence the configurational properties of strongly interacting matter by reshaping the energy-density profile along the holographic direction rather than merely modifying its integrated value. 
The simultaneous dependence on temperature, chemical
potential, and angular velocity motivates the
configurational characterization of distinct transition
regimes in the rotating QCD phase diagram.

At finite density, the bulk Abelian gauge sector
incorporates the dependence on the quark chemical
potential. Whereas several holographic implementations
employ the AdS-Reissner--Nordstr\"om approximation
\cite{Lee:2009bya,Colangelo:2010pe}, Andreev's exact
gauge-field solution \cite{Andreev:2010bv} retains
the exponential dependence on the soft-wall dilaton
scale. Its extension to rotating charged AdS geometries
\cite{Zhou:2021sdy,LEMOS199546,BravoGaete:2017dso,
Junqueira:2025xnx,Junqueira:2025evt}
provides the gravitational background adopted here.
The corresponding phase diagram exhibits first-order
confinement/deconfinement transitions and crossover-like
behavior in ensembles with distributed angular velocities.
Whether these regimes and the associated critical region
admit distinct configurational signatures beyond the
conventional HP analysis remains an open question. 
In this work, we investigate the CE of rotating strongly
interacting matter at finite density within Andreev's
soft-wall AdS/QCD model. Beyond the difference between
the competing on-shell actions, we obtain their individual
renormalized thermodynamic potentials by incorporating
the Gibbons--Hawking boundary contribution and
dilaton-dependent counterterms
\cite{Arutyunov:1998ve,Junqueira:2025evt}.
The bulk gauge-field regularity conditions
\cite{Horigome:2006xu,Nakamura:2006xk,Hawking:1995ap}
and its dilaton-dependent dynamics \cite{Braga:2015jca}
determine the finite-density contributions.
The resulting thermodynamic energies are used to construct energy-density profiles along the holographic coordinate and their modal
fractions for individual configurations at fixed angular
velocity and ensembles spanning a finite rotational
interval. Particular attention is devoted to the
configurational characterization of phase transitions
and the distinction between configurational and
thermodynamic stability near the critical region.

This paper is organized as follows.
Section~\ref{SecAndreev} introduces the rotating charged
gravitational backgrounds and their phase structure.
In Section~\ref{SecHR}, we implement the holographic
renormalization procedure and derive the corresponding
grand-canonical thermodynamic quantities.
Section~\ref{SecCE} develops the configurational
description of the confined and deconfined phases,
examining the behavior of the CE across the
confinement/deconfinement transition.
Section~\ref{SecGSI} revisits the Gleiser--Stamatopoulos
stability criterion in light of the thermodynamic
response functions and discusses the emergence of
CE extrema near the critical
region of the holographic QCD phase diagram.
Finally, we present our conclusions and perspectives.

\section{Andreev's AdS/QCD soft wall model for a rotating strongly-interacting matter}
\label{SecAndreev}

We adopt the rotating charged soft-wall AdS/QCD
background developed in Refs.~\cite{Junqueira:2025xnx,
Junqueira:2025evt}, summarizing here the ingredients
required for the subsequent holographic renormalization
and CE analysis. Unlike the
determination of the confinement/deconfinement phase
diagram, which relies on the difference between the
competing on-shell actions, the present formulation
requires the individual thermodynamic energies of
both gravitational configurations.

Within the AdS/QCD framework at finite density, a charged
BH carrying nonzero angular momentum in five-dimensional
AdS spacetime provides the gravitational description of
rotating strongly-interacting matter. Assuming cylindrical
symmetry, with matter undergoing uniform rotation at angular
velocity $\omega$ around a hypercylinder of radius $l$,
the corresponding BH geometry admits the canonical
representation \cite{Zhou:2021sdy, LEMOS199546, BravoGaete:2017dso}:
\begin{equation}\label{canon}
ds^2=-N(z,q)\,dt^2
+\frac{L^2}{z^2f(z,q)}\,dz^2
+R(z,q)\left[d\phi+P(z,q)dt\right]^2
+\frac{L^2}{z^2}\sum_{i=1}^{2}dx_i^2,
\end{equation}
where the metric functions are given by
\begin{eqnarray}
	N(z,q) &=& \frac{L^2}{z^2} \frac{(1-\omega^2 l^2) f(z,q)}{1- \omega^2 l^2 f(z,q)}\;, \\
	R(z,q) &=& \frac{L^2\,l^2\,\gamma^2}{z^2}\left(1  -  f(z,q) \omega^2 l^2\right)\;, \\
	P(z,q) &=& \frac{\omega(1-f(z,q))}{1- \omega^2 l^2f(z,q)}\;,
\end{eqnarray}
where $\gamma = 1/\sqrt{1 - l^2 \omega^2 }$ denotes
the Lorentz factor associated with the rotational motion,
and the blackening function reads
\begin{equation}\label{f(z)}
f(z,q) = 1 - \frac{z^4}{z_h^4}-q^2\, z_h^2 z^4 + q^2 z^6\;,
\end{equation}
with the horizon coordinate $z_h$ determined by
$f(z_h,q) = 0$. The parameter $q$ encodes the BH charge
($Q$). The geometry \eqref{canon} holographically
represents the deconfined QGP phase, whereas its thermal
AdS counterpart, recovered upon taking
$f(z,q) \rightarrow 1$, describes confined hadronic
matter.

The temperature associated with the deconfined phase
follows from the surface gravity of the rotating charged
BH. The Hawking temperature reads \cite{Zhou:2021sdy}
\begin{align}\label{HTrot}
T(q,\omega)
&=\frac{|\kappa_G|}{2\pi}
=\frac{1}{4\pi}\lim_{z\to z_h}
\left(\sqrt{\frac{g^{zz}}{N(z,q)}}
\left|\partial_z N(z,q)\right|
\right)
=\frac{2-q^2z_h^6}{2\pi\gamma z_h},
\end{align}
where $\kappa_G$ denotes the surface gravity and $g^{zz}$
is the component of the inverse metric along the holographic direction 
\eqref{canon}. Through the holographic dictionary, this
quantity is identified with the equilibrium temperature
of the QGP. Nonnegative temperatures restrict the
admissible horizon positions to
$z_h \leq \left(\sqrt{2}/q\right)^{1/3}$.
The corresponding Euclidean temporal period is given by 
$\beta = 1/T$ for the BH geometry, whereas for thermal
AdS one has
\begin{equation}\label{betaAdS}
\beta_{AdS} (q,\omega) = \beta(q, \omega) \sqrt{f(\epsilon, q)}\;,
\end{equation}
as follows from matching the asymptotic geometries at
the ultraviolet (UV) hypersurface $z=\epsilon$, in the limit
$\epsilon \to 0$. The temporal periods enter the
evaluation of the respective on-shell action densities,
which govern the HP transitions
\cite{Barbosa-Cendejas:2018mng, Junqueira:2025evt}
and provide the starting point for determining the
thermodynamic energies of the competing phases.

The gravitational backgrounds are constructed within
Einstein gravity with a negative cosmological constant
$\Uplambda = - \frac{12}{L^2}$, where $L$ is the AdS
curvature radius and $R = - \frac{20}{L^2}$ is the
Ricci scalar of pure AdS spacetime. The Euclidean
gravitational action underlying the five-dimensional
soft-wall prescription is expressed as
\cite{Karch:2006pv, Herzog:2006ra,BallonBayona:2007vp}
\begin{equation}\label{action1}
	I_G = - \frac{1}{ 2 \upkappa^2} \int_0^{z_h} dz\int d^4x \sqrt{g} e^{-\Phi}\left( R - \Uplambda \right) \;,
\end{equation}
where $\upkappa$ is the gravitational coupling constant.
The quadratic dilaton background $\Phi(z) = cz^2$
implements conformal symmetry breaking by introducing
the characteristic IR  mass scale $\sqrt{c}$.
The metric determinant assumes the same form for both
gravitational configurations, namely,
$g = l^2 L^{10}/z^{10}$. Within the adopted soft-wall
prescription, the gravitational contribution to the
on-shell action is written as
\begin{equation}\label{action1onshell}
I_{G_{\textsc{on-shell}}} = \frac{4l^2L^3}{\upkappa^2} V_{3D} \int_0^{\beta_s} dt \int_0^{z_h} dz\, \frac{e^{-cz^2}}{z^5}\;\,,
\end{equation} 
where $\beta_s$ specifies the Euclidean temporal period
of each geometry and $V_{3D}$ accounts for the spatial
volume factor. In the absence of an event horizon,
as occurs for thermal AdS, the integration along the holographic coordinate extends to infinity, corresponding to the prescription
$z_h \rightarrow \infty$.

Finite-density effects are incorporated through a bulk
Abelian vector field $V_\mu$, which accounts for the
BH charge and its holographic connection with the
density of strongly-interacting matter. The dynamics
of this sector in the presence of the dilaton background
is governed by the five-dimensional action
\cite{Braga:2015jca}
\begin{equation}\label{actionVF}
    I_{\textsc{VF}} = -\frac{1}{4g_5^2}\int_0^{z_h} dz \int d^4x \sqrt{g} e^{-\Phi} F_{MN}F^{MN}\;,
\end{equation}
with $F_{MN} = \partial_{[M} A_{N]}$ denoting the
associated field strength. According to the holographic
dictionary, the boundary value of the temporal component
$A_0$ encodes the quark chemical potential ($\mu$),
acting as the source for the density operator of the
dual gauge theory. The corresponding quark number density is associated
with the temporal component of the quark current,
$J^0=\bar{\psi}\gamma^0\psi$, with an implicit sum
over color and flavor indices.

Retaining the full dilaton dependence of the gauge
sector, rather than resorting to its near-boundary
approximation, Andreev's exact gauge-field solution
for the nonrotating configuration reads
\cite{Andreev:2010bv, Junqueira:2025xnx}
\begin{eqnarray}\label{A0final}
    A_0^{\textsc{nrs}}(z) = i\mu \left( \frac{e^{c z_h^2} - e^{c z^2}}{e^{c z_h^2}-1}\right).
\end{eqnarray}
The corresponding gauge-field components in the
presence of rotation are given by
\cite{Junqueira:2025xnx, Junqueira:2025evt}
\begin{eqnarray}
    A_0 &=& \gamma(\omega l) A_0^{\textsc{nrs}} \;,\label{A0}\\
    A_\phi & = &-l^2\omega \gamma(\omega l) A_0^{\textsc{nrs}},\label{Aphi}\\
    A_{x_1} &=& A_{x_2} = A_z = 0\;, \label{AiAz}
\end{eqnarray}
(using the nonrotating reference system acronym NRS), satisfying the horizon boundary condition
$A_0(z_h) = 0$. The latter is consistent with the
regularity requirement on the gauge-field norm
\cite{Horigome:2006xu, Nakamura:2006xk, Hawking:1995ap, Ballon-Bayona:2020xls}
and establishes the connection between the charge
parameter $q$, or equivalently the BH charge $Q$,
and the quark chemical potential
\cite{Lee:2009bya},
\begin{eqnarray}\label{qmu}
    \frac{Q}{c} = \frac{\eta q}{c} = \frac{\mu}{e^{c z_h^2}-1}\;,  
\end{eqnarray}
where the proportionality coefficient is defined as
\begin{eqnarray}
    \eta = \sqrt{\frac{3 g^2_5 L^2}{2 \upkappa^2}}\;.
\end{eqnarray}
Then, using Eqs. \eqref{A0} - \eqref{AiAz}, the relation \eqref{qmu} and assuming $\eta = 1$, one obtains the   following on-shell $U(1)$ action for the rotating system: 
\begin{eqnarray}\label{action2}
    I_{\textsc{VF}_{\textsc{on-shell}}} =  \frac{2lLc^2 \mu^2}{g^2_5(e^{cz_h^2}-1)^2 }\gamma^4 V_{3D} \, \int_0^{\beta_s} dt  \int_\epsilon^{z_\textsc{min}}dz\, z e^{-c z^2}\left[(1+l^2\omega^2)^2 + 4l^2\omega^2f(z,\mu)\right]\;,
\end{eqnarray}
which, together with \eqref{action1}, defines the total on-shell action
\begin{equation}\label{totalI}
    I_{\textsc{on-shell}} = I_{G_{\textsc{on-shell}}} + I_{\textsc{VF}_{\textsc{on-shell}}}\;. 
\end{equation}
Defining the action density by $\mathcal{E} = \frac{1}{lV_{3D}} I_{\textsc{on-shell}}$, Eq. \eqref{totalI} yields
\begin{equation}\label{Es}
	\mathcal{E}_s(\varepsilon) =  \beta_s  \int_\varepsilon^{z_\textsc{min}} dz\, \frac{e^{-c z^2}}{z^5}\left[ \frac{4L^3}{\upkappa^2} + \frac{2Lc^2\mu^2}{g^2_5(e^{cz_h^2}-1)^2}\gamma^4\left((1+l^2\omega^2)^2+4l^2\omega^2f(z,\mu)\right) z^6\right]\;, 
\end{equation}
wherein $\varepsilon$ is the UV  regulator of the integration over $z$. The deconfinement temperature $T_c(\mu,\omega)$ for a single state is computed when 
\begin{equation}\label{HPtransition}
    \bigtriangleup \bar{\mathcal{E}}(\bar{\mu},
 \omega l, \bar{z}_h)=\lim_{\varepsilon \rightarrow 0} \left[\mathcal{E}_{BH} - \mathcal{E}_{AdS} \right] = 0, 
\end{equation} 
at  $\bar{z}_h = \bar{z}_{h_c}(\bar{\mu}, \omega l)$. Eq. 
\eqref{HPtransition} is free of UV divergences, with $T_c(\mu,\omega)$ describing the transitions of the 3D-QCD phase diagram   \cite{Lee:2009bya, Braga:2022yfe, Braga:2025eiz}. In order to compute the total energy of the system in each phase, we must eliminate the UV divergences present in $\mathcal{E}_{BH}$ and $\mathcal{E}_{AdS}$ individually, which can be done through the introduction of the Gibbons--Hawking (GH) surface term into the total action  \cite{Arutyunov:1998ve, BallonBayona:2007vp}.  

\section{Holographic renormalization}
\label{SecHR}

 The holographic renormalization consists of introducing a boundary GH term into the action, with its corresponding surface counterterms. In the AdS/QCD approach, this allows us to construct renormalized Gibbs free energies, without UV divergences, in both the hadronic and plasma phases. In the soft-wall AdS model, see \cite{Arutyunov:1998ve, BallonBayona:2007vp}, the GH action is given by
\begin{equation}
I_{GH} = -  \frac{1}{\kappa^2} \int_{\partial M} d^4x \sqrt{h}\,  u(\Phi) K\;, 
\end{equation}
where $u(\Phi)$ is a function of the dilaton background field,  $K$ is the trace of the extrinsic curvature of the boundary, and $h$ is the determinant of the boundary induced metric $h_{\mu\nu}$, such that 
\begin{equation}
K = \nabla n^a = \frac{1}{\sqrt{g}} \partial_a \left( \sqrt{g} \,n^a \right)\;,
\end{equation}
where $n^a$ is a unitary vector normal to the boundary. The corresponding surface counterterm action takes the form
\begin{equation}
I_{ct} = \frac{1}{\kappa^2} \int_{\partial M} \sqrt{h} \,u(\Phi) F(L, \tilde{R}, \nabla \tilde{R})\;,
\end{equation}
where $F$ is a finite series of diffeomorphism invariants constructed from the AdS radius $L$
and the boundary induced curvature $\tilde{R}$. The total surface action density is given by
\begin{equation}\label{Isurf}
\mathcal{E}_{\text{surface}} = \mathcal{E}_{GH}+\mathcal{E}_{ct}\;,
\end{equation}
where $\mathcal{E}_{GH} = I_{GH}/lV_{3D}$ ($\mathcal{E}_{ct} = I_{ct}/lV_{3D}$). 
The functions $u$ and $F$ are fixed by the terms that eliminate the UV divergences present in the bulk action \eqref{Es}, and are given by
\begin{equation}\label{uF}
u = -1 + 2\Phi + \Phi^2 \text{log}\,\Phi \quad \text{and} \quad F = \frac{3}{L}\;.
\end{equation}
For a detailed explanation of this result, see \cite{BallonBayona:2007vp, Junqueira:2025evt}.

\subsection{Renormalized actions in the hadronic and QGP phases}
\label{SecActions}
Combining the bulk contribution with the boundary terms
specified above, the renormalized action density for each
gravitational configuration reads
\begin{equation}\label{Ihr}
\mathcal{E}_{total}^{(hr)} = \mathcal{E}_{\text{bulk}} + \mathcal{E}_{\text{surface}}\;.
\end{equation}
Using \eqref{Isurf} and \eqref{uF} in each phase, one obtains
\begin{eqnarray}
\mathcal{E}_{\text{surface}}^{AdS} &=& \frac{L^3 \beta_{AdS}}{\kappa^2} \left[ -\frac{1}{\epsilon^4} + \frac{2c}{\epsilon^2} + c^2\text{log}(c\epsilon^2)\right] \;,\label{IsurfAdS}\\
\mathcal{E}_{\text{surface}}^{BH} &=& \frac{L^3 \beta}{\kappa^2} \left[ -\frac{1}{\epsilon^4} + \frac{2c}{\epsilon^2} + c^2\text{log}(c\epsilon^2)+ \frac{1}{2 z_h^4}\right]\;,\label{IsurfBH}
\end{eqnarray}
such that, replacing \eqref{IsurfAdS} and \eqref{IsurfBH} into Eq. \eqref{Ihr}, and using the \eqref{Es} bulk actions for the corresponding geometries, the final renormalized action densities for the rotating hadronic and QGP matter, in their dimensionless versions, take the form:
\begin{eqnarray}\label{EhrAdS}
\!\!\!\!\!\!\!\!\!\!\bar{\mathcal{E}}_{AdS}^{(hr)} &=& \frac{L^3 \bar{\beta}}{2\kappa^2 \bar{z}_h^4} \left[(3-2\gamma_E)\bar{z}_h^4 + 3\bar{q}^2\gamma^4 \bar{z}_h^4 - 3\bar{q}^2l^2 \omega^2 \gamma^4(8-6\bar{z}_h^4-l^2\omega^2 \bar{z}_h^4)+ 24l^2\omega^2\bar{q}^4\gamma^4(3-\bar{z}_h^2)\right]\;,
\end{eqnarray}   
and 
\begin{eqnarray}\label{EhrBH}
\!\!\!\!\!\!\!\!\!\!\!\!\bar{\mathcal{E}}_{BH}^{(hr)} &=& \frac{L^3 \bar{\beta}}{2\kappa^2 \bar{z}_h^4} e^{-\bar{z}_h^2}\left[ 2(-1+\bar{z}_h^2)-3\bar{q}^2\bar{z}_h^4\gamma^4-3\bar{q}^2l^4\omega^4\bar{z}_h^4\gamma^4+6\bar{q}^2l^2\omega^2\gamma^4(4+4\bar{z}_h^2-\bar{z}_h^4)\right.\nonumber\\
&&-\left. 12l^2\omega^2\bar{q}^4\bar{z}_h^4\gamma^4(6+4\bar{z}_h^2+\bar{z}_h^4)+2e^{\bar{z}_h^2}+\bar{z}_h^4 e^{\bar{z}_h^2}(3-2\gamma_E+\bar{q}^2\bar{z}_h^2)+3\bar{q}^2\gamma^4 \bar{z}_h^4e^{\bar{z}_h^2}\right.\nonumber\\
&&-\left. 3\bar{q}^2l^2\omega^2\gamma^4e^{\bar{z}_h^2}(8-6\bar{z}_h^4-l^2\omega^2\bar{z}_h^4)-24\bar{q}^4l^2\omega^2\gamma^4 e^{\bar{z}_h^2}(3 - \bar{z}_h^2) + 2\bar{z}_h^4 e^{\bar{z}_h^2} \text{Ei}(-\bar{z}_h^2)\right]\;,
\,
\end{eqnarray}
where $\gamma_E$ is the Euler gamma constant, with the dimensionless variables defined as
\begin{eqnarray}
    \bar{z}_h &=& z_h \sqrt{c}\;,\nonumber\\
    \bar{\beta} &=& \frac{1}{\bar{T}} = \sqrt{c}\beta\;,\nonumber\\
    \bar{q} &=& q/c^{3/2}\;,\label{varsw}
    \end{eqnarray}
being $\bigtriangleup \bar{\mathcal{E}} = \upkappa^2 \bigtriangleup \mathcal{E}/(L^3 c^{3/2})$ the dimensionless action density. To obtain the physical values, we must use QCD phenomenology. In the soft wall holographic model, the IR parameter $\sqrt{c}$ can be fixed from the fit of the masses of the lightest $\rho$-mesons, which leads to $\sqrt{c}=338\, \text{MeV}$ \cite{Herzog:2006ra}. 

\subsection{Energy density in the grand-canonical ensemble}
\label{SecDensity}

To compute the total energy density in the grand-canonical ensemble, we need to obtain the Gibbs free energy ($G$), the entropy ($S$), and the averages of the total angular momentum ($J$) and total charge ($Q_e$) of the system. The dimensionless renormalized versions of these thermodynamic quantities are given by the expressions
\begin{eqnarray} \label{GSJQ}
   \bar{G} = \frac{1}{\beta} \bar{\mathcal{E}}^{(hr)}\;, \quad  \bar{S} = - \left(\frac{\partial \bar{G}}{\partial \bar{T}}\right)_{\bar{Q}}\;, \quad \bar{J} = - \left(\frac{\partial \bar{G}}{\partial \omega}\right)_{\bar{T},\bar{\mu}}\;, \quad \text{and} \quad  \bar{Q}_e = - \left(\frac{\partial \bar{G}}{\partial \bar{\mu}}\right)_{\bar{T},\omega}.
\end{eqnarray}
The natural thermodynamic variables of the grand-canonical
potential are the temperature, chemical potential, and
angular velocity. Accordingly, its differential reads
\begin{equation}\label{GrandCanonicalFirstLaw}
d\bar{G}
=-\bar{S}\,d\bar{T}
-\bar{Q}_e\,d\bar{\mu}
-\bar{J}\,d\omega.
\end{equation}
In the expression for $\bar{G}$ in (\ref{GSJQ}), we can use $\bar{\beta}_{AdS} = \bar{\beta}$, once $\bar{\mathcal{E}}^{(hr)}_{AdS}$ is free of UV divergences after applying holographic renormalization. For a single state at a given angular velocity, the Gibbs free energy for the confined and plasma phases is given by, respectively,
\begin{eqnarray}\label{Gs}
    \bar{G}_{AdS}(\bar{T}, \omega l, \bar{\mu}) = \frac{1}{\bar{\beta}} \bar{\mathcal{E}}^{(hr)}_{AdS}(\bar{T}, \omega l)\;,\quad \text{and} \quad  \bar{G}_{BH}(\bar{T}, \omega l, \bar{\mu}) = \frac{1}{\bar{\beta}} \bar{\mathcal{E}}^{(hr)}_{BH}(\bar{T}, \omega l)\;. 
\end{eqnarray}

To write the Gibbs energies as a function of $\bar{T}$, $\omega l$ and $\bar{\mu}$, we must find the horizon function $\bar{z}_h(T, \omega l, \bar{\mu})$. The sixth-order polynomial of the Hawking temperature \eqref{HTrot} in $\bar{z}_h$ does not have an analytical solution. To overcome this problem, we will expand $T(\bar{z}_h, \omega l, \bar{\mu})$ around a given $\bar{z}_{h0}$, in powers of $(\bar{z}_h - \bar{z}_{h 0})^n$ up to a given order that has an analytical solution. For the present numerical analysis, we will start the expansion from a point in the 3D phase diagram for the temperature $\bar{T}_0(\bar{\mu}_0, \omega_0$l), with $\bar{\mu}_0 =0.396$ and $\omega_0 l = 0.4$, where the dimensionless chemical potential is defined by 
\begin{eqnarray}\label{barmu}
    \bar{\mu} &=& \mu/\sqrt{c}\;.
\end{eqnarray}
In this case, $\bar{z}_{h0} = 0.986429 $, with the expansion valid inside the region where $\vert \bar{z}_h - \bar{z}_{h0}\vert < 1$. We will also choose $n=3$ to obtain the expression for $\bar{z}_h(\bar{T}, \omega l,\bar{\mu})$, which should be substituted into Eqs. (\ref{EhrAdS}, \ref{EhrBH}) to construct $\bar{G}_{AdS}$ and $\bar{G}_{BH}$ as functions of temperature and angular momentum.

By doing this, Eq. \eqref{Gs} is completely defined, and we are able to compute $\bar{S}$, $\bar{J}$ and $\bar{Q}_e$ as functions of $\bar{T}$ and $\omega l$, using Eq. \eqref{GSJQ}. Assuming a system described by a grand-canonical ensemble, the total energy of the system composed of a single state at a fixed angular momentum is then given by\footnote{It is implicit that these thermodynamic quantities are densities, divided by the radius $l$ and the spatial factor volume $V_{3D}$.}
\begin{eqnarray}\label{Es1}
E_s(\bar{T}, \omega l)= \bar{G} + \omega \bar{J} + \bar{\mu} \bar{Q}_e+ \bar{T} \bar{S}\;,   
\end{eqnarray}
which will be used to define the energy density per unit of $dz$ in each phase, $\rho(z, T, \omega l, \bar{\mu})$, such that
\begin{eqnarray}\label{Erho}
E_s(\bar{T}, \omega l, \bar{\mu})  = \int_0^{\bar{z}_f}\rho_s(\bar{z}, \bar{T}, \omega l, \bar{\mu})\, d\bar{z}   
\end{eqnarray}
wherein $\bar{z}_f = (\bar{z}_h, \infty)$ for the charged BH and thermal AdS geometries, respectively. This density, in turn, will define the modal fraction, which is written in terms of its absolute value in Fourier space, used to describe the CE of the rotating system. This CE will be valid for a single state. If we consider an ensemble with $N_p$ particles, the system will be composed of a combination of hadronic and QGP states with different angular momentum, which will make individual contributions to the total CE of the ensemble at a given temperature. Assuming that the angular momentum is described by discrete levels, one can write 
\begin{equation}\label{deltaw}
  (\omega l)_{\text{max}} - (\omega l)_{\text{min}} = N_p a_\omega\;,  \end{equation}
being $a_\omega$ the average interval between two neighboring levels of angular momentum, multiplied by $l$, in the large $N_p$ limit, one has $a_\omega \approx d(\omega l)$, such that the total energy for this type of ensemble will be expressed by
\begin{eqnarray}\label{GNp}
    E_{N_p}(\bar{T}, \bar{\mu}) =\frac{1}{a_\omega} \left[\int_{(\omega l)_{\text{min}}}^{ (\omega l)_{\text{max}}}  E_s(\bar{T}, \omega l,\bar{\mu})\, d(\omega l) \right]\;,
\end{eqnarray}
with the phase of the states varying along the angular momentum interval, according to relativistic rotation effects in the low-density regime, see \cite{Junqueira:2025evt}. As we shall see, the CE, via AdS/QCD duality, will serve as an indicator of the type of phase transition, ruled by the properties of strongly-interacting matter.

\section{Configurational entropy and phase transitions}
\label{SecCE}
The CE provides an information-theoretic 
measure of the spatial organization of localized physical 
configurations through their spectral decomposition \cite{Gleiser:2011di,Gleiser:2012tu}. Its 
construction is based on Shannon's information entropy, 
which quantifies the uncertainty associated with a discrete 
random variable characterized by a probability distribution 
$\{p_n\}$, and is defined as \cite{shannon}
\begin{equation}
 S_{\textsc{Shannon}} = - \sum_n \, p_n \log p_n \; .
 \label{discretepositionentropy}
\end{equation}
The extension of this information-theoretic framework to 
spatially continuous physical systems relies on the 
distribution of their Fourier modes, with the corresponding 
spectral weights playing the role of a continuous probability 
density \cite{Gleiser:2018kbq}. The AdS boundary has codimension one relative to the bulk, with the additional coordinate $z$ parametrizing the inverse energy scale of the dual gauge theory. Spatial homogeneity along the boundary restricts the spectral support of the bulk energy distributions to zero momentum in the boundary spatial directions. The nontrivial modal structure is therefore confined to the holographic direction, and the configurational entropy is evaluated in the corresponding one-dimensional Fourier space \cite{Bernardini:2025wkb,Bernardini:2018uuy,daRocha:2024sjn}. 
Therefore, for an effectively one-dimensional configuration, the CE is defined by the functional \cite{Gleiser:2018kbq}
\begin{equation}\label{SC}
 S_C[f] = - \int dk \, f({k}) \ln f({k}) \,,
\end{equation}
where $f(k)$ is the modal fraction, which characterizes the 
relative contribution of each Fourier mode to the spectral 
power of the underlying physical configuration. For a 
localized system described by an energy density $\rho$, 
the modal fraction is conventionally constructed from its 
Fourier transform, $\tilde{\rho}(k)$, according to
\begin{equation}
 f(k) = \frac{\vert \tilde{\rho}(k)\vert^2}{\mathcal{N}  } \,  
 \label{modalfraction}
\end{equation}
where the normalization factor $\mathcal{N}$ corresponds 
to the integrated spectral power over the entire momentum 
domain,
\begin{eqnarray}\label{N}
 \mathcal{N} = \int_{-\infty}^{+\infty} 
 \vert \tilde{\rho}(k)\vert^2\, dk\;.
\end{eqnarray}
By construction, the modal fraction is nonnegative and 
normalized to unity, thereby defining a probability density 
in momentum space. The CE thus probes how the spectral 
content of the energy density is distributed among its 
constituent Fourier modes, providing a quantitative 
characterization of the configurational information encoded 
in the underlying physical system. At a fixed momentum 
scale, configurations whose spectral power is concentrated 
within a narrower range of modes generally exhibit lower 
CE, whereas broader spectral distributions tend to yield 
higher values.

We emphasize that, in particular in the present case, the integration over the spatial bulk coordinates in the action density is trivial. It does not encode the information content or the dynamics of the system. If we have defined the energy density \eqref{Erho} in terms of integration over $d^4x$ instead of over the Poincaré coordinate, this would merely hide the contribution of a spatial volume factor. For this reason, we define the Fourier transform of the energy density as 
\begin{eqnarray}
    \tilde{\rho}_s(k, \bar{T}, \omega l,\bar{\mu}) = \frac{1}{2\pi}\int_{0}^{z_f} \rho_s(z, \bar{T}, \omega l,\bar{\mu})\,e^{ikz}\,dz\;,
\end{eqnarray}
which can be understood, in general, as a Fourier transform of the density up to a trivial factor without relation to the thermodynamics of the ensemble. The dependence of $ \tilde{\rho}_s$($ \tilde{\rho}_{BH}$ or $ \tilde{\rho}_{AdS}$) on the temperature, angular momentum, and quark chemical potential, for a single state, is directly inferred from Eqs. \eqref{Es} and \eqref{Erho}. In addition to that, to describe the modal fraction in a consistent way, the energy density in momentum space must obey two conditions:   
\begin{itemize}

    \item \textit{First condition (normalizability of the Fourier
    power spectrum)}: the squared modulus
    $\vert \tilde{\rho}(k,\bar{T},\omega l,\bar{\mu})\vert^2$
    must be integrable over the entire Fourier domain, ensuring
    that the normalization factor satisfies
    \begin{equation}
        0 < \mathcal{N}(\bar{T},\omega l,\bar{\mu})
        = \int_{-\infty}^{+\infty}
        \vert \tilde{\rho}(k,\bar{T},\omega l,\bar{\mu})
        \vert^2\,dk < \infty,
    \end{equation}
    or, equivalently, that $\mathcal{N}(\bar{T},\bar{\mu})$
    remains finite and nonvanishing for the ensemble.
    For localized and sufficiently regular energy distributions,
    the Fourier power spectrum is expected to exhibit the
    asymptotic behavior
    \begin{equation}
        \lim_{k\rightarrow\pm\infty}
        \vert\tilde{\rho}(k,\bar{T},\omega l,\bar{\mu})
        \vert^2=0.
    \end{equation}
    Although this asymptotic condition is not sufficient by
    itself to guarantee normalizability, the square integrability of the energy-density profile along the holographic direction, 
    $\rho\in L^2(\mathbb{R})$, provides a sufficient condition
    through Parseval's identity, provided that $\rho$ is not
    identically zero. Here, the radial distribution is understood
    to be extended by zero outside its corresponding holographic
    domain. Consequently, the modal fraction constitutes a
    properly normalized spectral probability density, satisfying
    \begin{equation}
        f(k,\bar{T},\omega l,\bar{\mu})\geq 0,
        \qquad
        \int_{-\infty}^{+\infty}
        f(k,\bar{T},\omega l,\bar{\mu})\,dk=1.
    \end{equation}

    \item \textit{Second condition (finiteness of the CE)}: in analogy with Shannon's information entropy,
    the modal fraction plays the role of a continuous probability
    density associated with the relative spectral weights of the
    Fourier modes. Accordingly, the CE
    density in Fourier space is defined as
    \begin{equation}\label{lambda}
        \lambda(k,\bar{T},\omega l,\bar{\mu})
        =
        -f(k,\bar{T},\omega l,\bar{\mu})
        \ln\left[f(k,\bar{T},\omega l,\bar{\mu})\right].
    \end{equation}
The existence of a finite
    CE requires
    \begin{equation}
        \int_{-\infty}^{+\infty}
        \left|\lambda(k,\bar{T},\omega l,\bar{\mu})\right|
        \,dk < \infty.
    \end{equation}
    For the localized energy distributions considered in the
    present framework, the entropy density is further expected
    to exhibit regular asymptotic behavior,
    \begin{equation}
        \lim_{k\rightarrow\pm\infty}
        \lambda(k,\bar{T},\omega l,\bar{\mu})=0.
    \end{equation}
    Whenever the modal fraction satisfies $0\leq f(k)\leq 1$,
    the corresponding entropy density is nonnegative throughout
    the Fourier domain. Nevertheless, unlike discrete Shannon
    entropy, differential entropy does not generally require
    pointwise positivity of its entropy density, since a
    normalized continuous probability density may locally
    exceed unity. The positivity of $\lambda(k)$ should
    therefore be regarded as an additional property of the
    particular spectral distributions satisfying this bound,
    rather than a universal mathematical requirement for the
    existence of CE. Throughout this work,
    the Fourier variable $k$ is conjugate to the dimensionless
    holographic coordinate $\bar{z}=\sqrt{c}\,z$, thereby fixing
    the reference scale entering the differential entropy
    functional.

\end{itemize}
These two conditions are required for the CE analysis according to the Gleiser--Stamatopoulos interpretation concerning the stability of physical systems \cite{Gleiser:2011di,Gleiser:2012tu,Gleiser:2013mga}. This will allow us to associate the behavior of the CE, through gauge/gravity duality, with the dynamics and degrees of freedom of the system, in the confined and deconfined phases, determining how the CE responds to the plasma vorticity, and how it can be correlated to the QCD phase diagram. For this, we need to calculate the CE of the rotating matter in the charged BH and thermal AdS geometries, for different system configurations in the soft-wall AdS/QCD model. Our goal is to find a correlation between the CE and the type of confinement/deconfinement phase transitions in different density regimes (first-order transitions or crossovers), and to compare the thermodynamic instability (energy loss through photon emissions) with the instability described by the CE.  

\subsection{First-order phase transitions}
\label{Sec1st}
For a single state at a given angular velocity, the energy density must be a solution of Eq. \eqref{Erho}, with $E_s(\bar{T}, \omega l)$, see Eq. \eqref{Es}, defined by Eqs.  (\ref{EhrAdS}, \ref{EhrBH}) for the hadronic and plasma phases, respectively, with the renormalized Gibbs free energies given by Eq. \eqref{Gs}. The other thermodynamic quantities ($S$, $J$ and $Q_e$) that appear in the grand-canonical total energy are obtained from it, using Eq. \eqref{GSJQ}. In addition to obeying these equations, the modal fraction solution must obey the two conditions outlined in the previous section. (To facilitate graphical analysis, we will consider variations in temperature and angular momentum at $\bar{\mu}_0 = 0.396$.) According to this, the solutions in each phase can be given by the following Gaussian energy distributions in the Poincaré direction with different amplitudes\footnote{It is implicit that $\rho_s(z, \bar{T}, \omega l) = \rho_s(z, \bar{T}, \omega l, \bar{\mu} = \bar{\mu}_0)$; analogously for $E_s$.}:
\begin{eqnarray}
    \rho_{AdS}(\bar{z}, \bar{T}, \omega l) &=& \frac{2}{\sqrt{\pi}} \,E_{AdS}( \bar{T}, \omega l)\,e^{-\bar{z}^2}\;,\label{rhoAdS}\\
    \rho_{BH}(\bar{z}, \bar{T}, \omega l)&=&\frac{2}{\sqrt{\pi}}\frac{1}{\text{Erf}(\bar{z}_h)}E_{BH}( \bar{T}, \omega l)\,e^{-\bar{z}^2}\;.\label{rhoBH}
\end{eqnarray}
where $\text{Erf}(t) = \frac{2}{\sqrt{\pi}}\int_0^t e^{-y^2}\,dy$ is the error function. These Gaussian distributions are not unique solutions, since
the integrated energy $E_s$, for each gravitational background,
is invariant under the transformation
\begin{eqnarray}\label{rhoTransformation}
    \rho_s(\bar{z},\bar{T},\omega l)
    \mapsto
    \rho_s(\bar{z},\bar{T},\omega l)
    +\partial_{\bar{z}}\sigma_s(\bar{z},\bar{T},\omega l)\;,
\end{eqnarray}
where $\sigma_s$ denotes an arbitrary sufficiently regular
function whose derivative with respect to the holographic coordinate yields a surface contribution to the integrated energy, admitting the boundary conditions
$\displaystyle\lim_{\bar{z}\to\infty}\sigma_{AdS}(\bar{z})=0=\lim_{\bar{z}\to\bar{z}_h}\sigma_{BH}(\bar{z})$, supplemented by
$\displaystyle\lim_{\bar{z}\to0}\sigma_s(\bar{z})=0$ at the UV  boundary.
Indeed, the corresponding variation of the integrated energy
reads
\begin{eqnarray}
    \delta E_s
    &=&
    \int_0^{\bar{z}_f}
    \partial_{\bar{z}}\sigma_s
    (\bar{z},\bar{T},\omega l)\,d\bar{z}
    =
    \sigma_s(\bar{z}_f,\bar{T},\omega l)
    -\sigma_s(0,\bar{T},\omega l)=0\;,
\end{eqnarray}
where $\bar{z}_f\to\infty$ for thermal AdS and
$\bar{z}_f=\bar{z}_h$ for the BH geometry.
More generally, the invariance of $E_s$ only requires
$\sigma_s(\bar{z}_f)=\sigma_s(0)$, with the vanishing of
both boundary contributions constituting a sufficient
condition.

Although the transformation \eqref{rhoTransformation}
preserves the integrated thermodynamic energy, it does not, in general, preserve the spectral content of the energy-density profile along the holographic direction. Under this transformation, the
corresponding Fourier amplitude changes according to
\begin{eqnarray}
    \delta\tilde{\rho}_s(k)
    &=&
    \frac{1}{2\pi}
    \int_0^{\bar{z}_f}
    \partial_{\bar{z}}\sigma_s(\bar{z})
    e^{ik\bar{z}}\,d\bar{z}
  =
    -ik\tilde{\sigma}_s(k)\;,
    \label{FourierSurfaceTransformation}
\end{eqnarray}
where the vanishing boundary conditions imposed above
have been employed and $\tilde{\sigma}_s(k)$ denotes
the Fourier transform of $\sigma_s(\bar{z})$.
The resulting Fourier power spectrum therefore transforms as
\begin{eqnarray}
    |\tilde{\rho}_s(k)|^2
    &\mapsto&
    |\tilde{\rho}_s(k)|^2
    +k^2|\tilde{\sigma}_s(k)|^2
+
    2k\,\operatorname{Im}
    \left[\tilde{\rho}_s^{*}(k)
    \tilde{\sigma}_s(k)\right]\;.
    \label{PowerSpectrumTransformation}
\end{eqnarray}
In particular, the zero Fourier mode remains invariant,
whereas the nonzero modes may undergo a redistribution
of their spectral weights. Consequently, radial energy
distributions that are equivalent at the level of integrated
thermodynamic quantities need not be configurationally
equivalent.

The Gaussian distributions adopted here constitute a
particular representative of this class of energetically
equivalent radial profiles, corresponding to the prescription
$\partial_{\bar{z}}\sigma_s=0$ relative to the Gaussian ansatz.
Their functional dependence is motivated by the quadratic
dilaton background underlying the soft-wall model,
$e^{-\Phi(z)}=e^{-\bar{z}^2}$, whereas their normalization
is determined by the corresponding thermodynamic energies.
Moreover, their distinct radial supports incorporate the
geometric differences between the confined and deconfined
phases. 
This prescription establishes a common functional basis
for comparing the configurational properties of both
gravitational backgrounds, without introducing additional
profile parameters beyond those already present in the
soft-wall construction. The resulting CE thus probes the spectral structure of the prescribed holographic energy-density profiles, capturing the interplay between the characteristic soft-wall scale and the extent of the holographic domain of the corresponding gravitational configurations. 
The possible dependence of its quantitative features on
alternative energetically equivalent profiles should,
nevertheless, be distinguished from those configurational
properties that persist under admissible deformations
of the radial energy distribution. 

In Fourier space for the confined phase, one has   \begin{eqnarray}\label{tilderhoAdS}
        \tilde{\rho}_{AdS}(k, \bar{T}, \omega l) &=& \frac{1}{\pi^{3/2}}\left( \frac{\sqrt{\pi}}{2}e^{-k^2/4}+i\mathcal{D}_+\left(\frac{k}{2}\right)\right)E_{AdS}( \bar{T}, \omega l)\;,
\end{eqnarray} 
where $\mathcal{D}_+(t) = e^{-t^2}\int_0^t dy\,e^{y^2}$ is the Dawson integral, such that
\begin{eqnarray}\label{AbstilderhoAdS}
     \vert\tilde{\rho}_{AdS}(k, \bar{T}, \omega l)\vert^2 = \frac{1}{\pi^3}\left[ \frac{\pi}{4}e^{-k^2/2}+\left(\mathcal{D}_+(\frac{k}{2})\right)^2\right]E^2_{AdS}( \bar{T}, \omega l)\;.
\end{eqnarray}
An immediate consequence of the adopted radial prescription
is the invariance of the CE under
overall energy rescalings. More generally, for a radial
energy distribution admitting the factorization
\begin{equation}\label{GeneralFactorization}
    \rho_s(\bar{z};\boldsymbol{\xi})
    =E_s(\boldsymbol{\xi})g_s(\bar{z};\boldsymbol{\xi})\;,
\end{equation}
where $\boldsymbol{\xi}=(\bar{T},\bar{\mu},\omega l)$
and $\int_0^{\bar{z}_f}g_s(\bar{z};\boldsymbol{\xi})
\,d\bar{z}=1$, the modal fraction, provided that
$E_s\neq0$, reduces to
\begin{equation}\label{AmplitudeInvariantModal}
    f_s(k;\boldsymbol{\xi})
    =
    \frac{|\tilde{g}_s(k;\boldsymbol{\xi})|^2}
    {\displaystyle\int_{-\infty}^{+\infty}
    |\tilde{g}_s(k';\boldsymbol{\xi})|^2\,dk'}\;.
\end{equation}
The thermodynamic dependence of the CE is therefore governed by changes in the shape
and radial support of the normalized energy distribution,
rather than by its overall energy amplitude.
This property is not restricted to Gaussian distributions.
In particular, the thermal AdS prescription
\eqref{rhoAdS} yields
\begin{equation}\label{ConstantCETheorem}
    \partial_{\bar{T}}S_C^{AdS}
    =
    \partial_{\bar{\mu}}S_C^{AdS}
    =
    \partial_{\omega l}S_C^{AdS}
    =0\;,
\end{equation}
whereas the finite radial support of the BH
distribution \eqref{rhoBH} introduces an implicit
thermodynamic dependence through the horizon coordinate
$\bar{z}_h(\bar{T},\bar{\mu},\omega l)$.

\begin{figure}[!htb]
	\centering
    \includegraphics[scale=0.5]{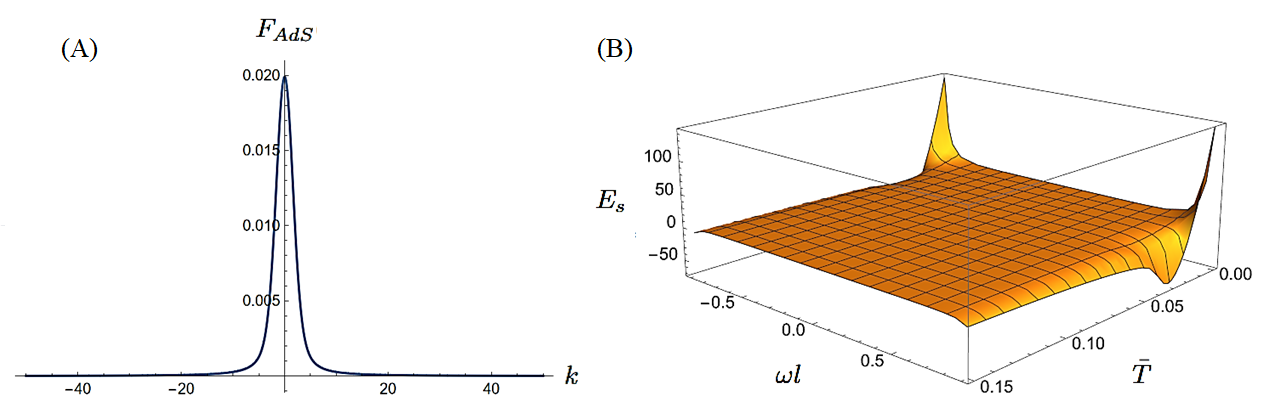}
	\caption{(A) Function $F_{AdS}(k)$, and (B) total energy in the grand-canonical ensemble, for a strongly interacting matter in the confined phase at  $\bar{\mu} = 0.396$, as a function of the temperature ($\bar{T}$) and the rotational velocity ($\omega l$).}
    \label{FEs}
\end{figure}

In Fig. \ref{FEs}, we have plot the function 
\begin{eqnarray}\label{FAdS}
    F_{AdS}(k) = \frac{\vert\tilde{\rho}_{AdS}(k, \bar{T}, \omega l)\vert^2}{E^2_{AdS}(\bar{T}, \omega l)}
\end{eqnarray}
and the total total energy $E_{AdS}( \bar{T}, \omega l)$. The curve in Fig. \ref{FEs}-(A) shows that the solution \eqref{tilderhoAdS} obeys the \textit{first condition} for the modal fraction. The surface $E_{AdS}( \bar{T}, \omega l)$, see Fig. \ref{FEs}-(B), does not have divergencies to compete with $F(k)$, so that  $\vert\tilde{\rho}_{AdS}(k \rightarrow\pm\infty, \bar{T}, \omega l)\vert^2 \rightarrow 0$, with $ \vert\tilde{\rho}_{AdS}(k, \bar{T}, \omega l)\vert^2$ finite along
the entire $k$-axis. 

\begin{figure}[!htb]
	\centering
    \includegraphics[scale=0.5]{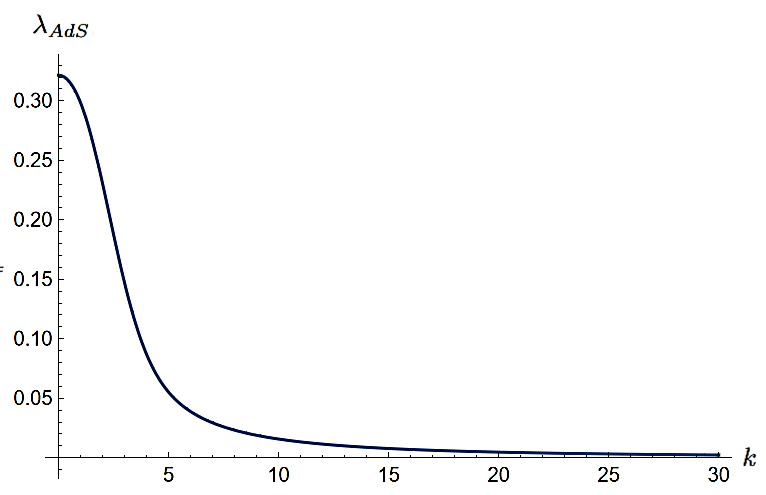}
	\caption{Function $\lambda_{AdS}(k)$, representing the CE density per unit of momentum, for the matter in the hadronic confined phase.}
    \label{lambdaAdS}
\end{figure}

Replacing  \eqref{AbstilderhoAdS} into \eqref{N}, one obtains
\begin{eqnarray}\label{NAdS}
    \mathcal{N}_{AdS}(\bar{T}, \omega) = \frac{1}{\sqrt{2}}E^2_{AdS}( \bar{T}, \omega l)\;. 
\end{eqnarray}
This result for the normalization shows that $ \lambda_{AdS}(k)$, defined in Eq. \eqref{lambda}, obeys the \textit{second condition}, see Fig. \ref{lambdaAdS}, as this curve is independent of the temperature and rotation of the hadrons. We plotted the function only for positive $k$, since it is even, $\lambda(k) = \lambda(-k)$. One observes that  $ \lambda_{AdS}(k)$ is always positive, satisfying $ \lambda_{AdS}(k\rightarrow\pm \infty) \rightarrow 0$. It is easy to verify that the same behavior appears at any value of the quark chemical potential. Consequently, using Eq. \eqref{SC}, the CE of a single state in the confined phase is constant, given by
\begin{eqnarray}\label{CEAdS}
    S_{CAdS} = 2.60633\;. 
\end{eqnarray}

Now, taking the Fourier transform of \eqref{rhoBH}, one obtains the following energy density for QGP states in momentum space:
\begin{eqnarray}\label{tilderhoBH}
\tilde{\rho}_{BH}(k, \bar{T}, \omega l) =\frac{e^{-k^2/4}}{2\pi\text{Erf}(\bar{z}_h)}\left[\text{Erf}\left(\bar{z}_h - i\frac{k}{2}\right)+i\text{Erfi}\left(\frac{k}{2}\right)\right]E_{BH}( \bar{T}, \omega l)\;,
\end{eqnarray}
where $\text{Erfi}(t) = -i\text{Erf}(it)$ is the imaginary error function. From the expression above, one concludes that BH CE will depend on the temperature and $\omega l$ (we are considering $\bar{\mu} = \bar{\mu}_0$), as the BH normalization cannot be written in the form \eqref{NAdS}, with $\mathcal{N}_{BH}$ given by $E_{BH}^2$ multiplied by a constant. The absolute value of \eqref{tilderhoBH} can be written as
\begin{equation}
 \vert\tilde{\rho_{BH}}(\bar{k}, \bar{T}, \omega l)\vert^2=\frac{1}{\pi^3}\left(\frac{E_{BH}( \bar{T}, \omega l)}{\text{Erf}(\bar{z}_h)}\right)^2\left[\left(\int_0^{\bar{z}_h}d\bar{z}\,e^{-\bar{z}^2}\cos(k\bar{z})\right)^2+\left(\int_0^{\bar{z}_h}d\bar{z}\,e^{-\bar{z}^2}\sin(k\bar{z})\right)^2\right]\;.
\end{equation}

\begin{figure}[!htb]
	\centering
    \includegraphics[scale=0.54]{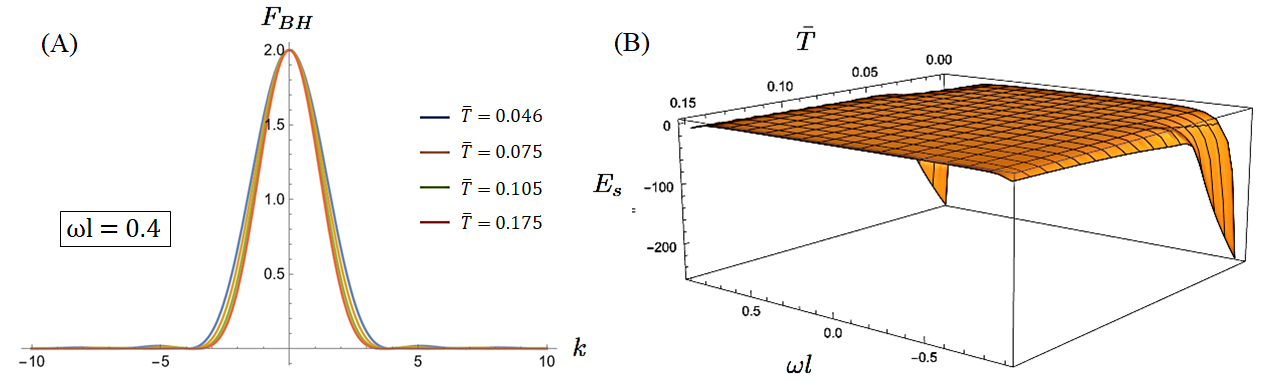}
	\caption{(A) Function $F_{BH}(k)$  at fixed rotational velocity ($\omega l = 0.4$) and different temperatures, and (B) total energy in the grand-canonical ensemble in the deconfined phase at  $\bar{\mu} = 0.396$, as a function of $\bar{T}$ and $\omega l$.}
    \label{FBHEs}
\end{figure}

\begin{figure}[!htb]
	\centering
    \includegraphics[scale=0.5]{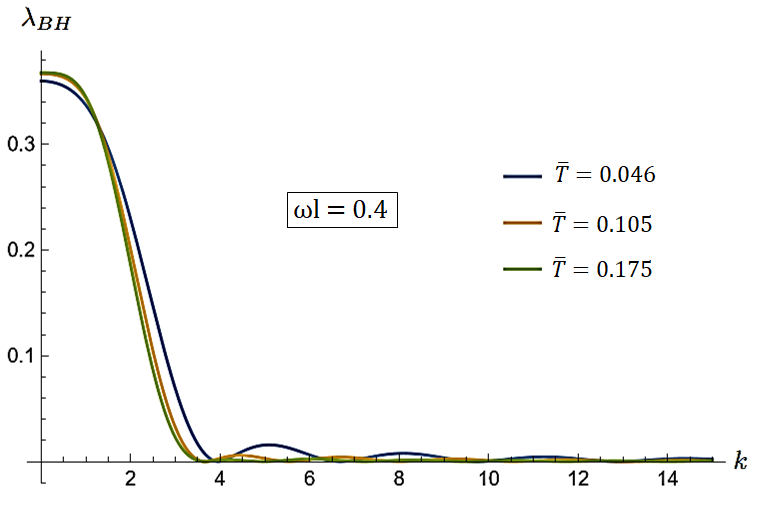}
	\caption{Function $\lambda_{BH}(k)$ for the matter in the deconfined phase, at a fixed rotational velocity ($\omega l = 0.4$) and different temperatures.}
    \label{lambdaBH}
\end{figure}

In Fig. \ref{FBHEs} we have plot $E_{BH}$ and the function
\begin{eqnarray}
 F_{BH}(k, \bar{T}, \omega l) = \frac{\vert\tilde{\rho}_{BH}(k, \bar{T}, \omega l)\vert^2}{E^2_{BH}( \bar{T}, \omega l)}\;,
\end{eqnarray}
for a single state at a fixed rotational velocity ($\omega l = 0.4$), varying the temperature in the phase diagram along the chemical potential $\bar{\mu} = \bar{\mu}_0$. As we can see, $F_{BH}$ depends on the temperature and $\omega l$ (it is easy to verify that a similar behavior is observed at any rotational velocity). For the same reasons discussed for the thermal space, analyzing the curves and the surface of figures $(A)$ and $(B)$ below, we conclude that the BH density, given by the solution \eqref{tilderhoBH} in momentum space, satisfies the \textit{first condition} for the modal fraction. The corresponding $\lambda_{BH}(k, \bar{z}_h, \omega l)$ is even and was plotted in Fig. \ref{lambdaBH} at different temperatures, from which one concludes that it also satisfies the \textit{second condition}. 

\begin{figure}[!htb]
	\centering
    \includegraphics[scale=0.54]{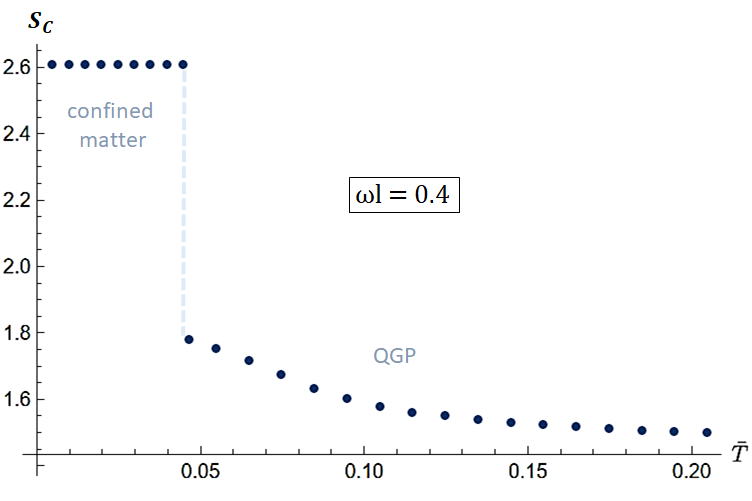}
	\caption{\textit{First-order phase transition.} CE of a single state at $\omega l  = 0.4$, as a function of the temperature. }
    \label{SC1st}
\end{figure}

To compute the QGP modal fraction normalizations, we need to integrate \eqref{tilderhoBH} over momentum. As it does not have an analytical solution, we are forced to employ numerical methods. The values of $\mathcal{N}_{BH}$ for a single state in the plasma phase at different temperatures varying along $\bar{\mu} = \bar{\mu}_0$, rotating with $\omega l = 0.4$, were displayed in Table \ref{table1}, see Appendix \ref{app1}. Using Table \ref{table1} and applying the numerical analysis to Eq. \eqref{SC}, one obtains the CE curve in the deconfined phase, described by the points in Fig. \ref{SC1st}. As we can see, there is a gap at the critical temperature $\bar{T}_c(\bar{\mu}_0,0.4) = \bar{T}_0 = 0.0466958$, which is computed using the HP transition equation  \eqref{HPtransition}. This gap characterizes the confinement/deconfinement as a first-order transition, indicating the jump of degrees of freedom from $N_c^0$ (confined matter) to $N_c^2$ (QGP phase), see \cite{BallonBayona:2007vp}. The CE of the hadronic matter is constant, reinforcing the interpretation of  constant CE as an indicator of color confinement, where the informational content of the state has no intrinsic dynamics. A similar result was obtained for nonrotating matter in the hard wall AdS/QCD model at zero density, see \cite{Braga:2020opg}. The difference is the value of the constant, which showed to be sensitive to way we introduced the IR scale into the theory (the result \eqref{CEAdS} is also valid for $\omega l =  0$ and $\bar{\mu} = 0$).

The statement that the vanishing of the thermodynamic entropy serves as an indicator for color confinement, developed in \cite{BallonBayona:2007vp} for nonrotating matter in the hard wall model, seems to be weaker if compared to the constancy of the CE in the confined phase. For rotating matter in the exact soft wall case, the hadronic entropy is not constant but increases with the temperature; see Fig. 9 of \cite{Junqueira:2025evt}, indicating that $S_{AdS}$ contains information involving other dynamics of the system, associated with the rotation (angular momentum) of the particles. In this sense, a constant CE appears as a stronger indicator of color confinement, since the result \eqref{CEAdS} proved to be independent of the way in which matter rotates and also of its density. It suggests a correlation between the CE and color dynamics, with the information ``confined" (constant CE) in the hadronic phase. First-order transitions appear in the high-density regime in the QCD phase diagram. These transitions are dominated by the curve of nonrotating matter, that is, for states rotating at a fixed rotational speed, and are thus consistently described by the transition indicated by the CE, according to the gap in Fig. \ref{SC1st}.

The relation between the CE of a physical state and its thermodynamics appears to be more subtle than that proposed in \cite{Braga:2020opg}. According to Gleiser--Stamatopoulos interpretation, the instability of a physical system increases as the CE increases. From Fig. \ref{SC1st}, one can conclude that the plasma becomes more stable as the temperature increases, since the BH CE decreases with $\bar{T}$. Apparently, the CE correlates well with thermodynamic stability, as the specific heat is always positive for a single state rotating with $\omega l = 0.4 $, see Fig. \ref{CvFig}-(A), describing a gain in mass of the system. However, there is no CE response to variations in the specific heat as the temperature increases. This leads us to believe that there is no direct correlation between thermodynamic stability and CE. Certainly, this does not occur in the confined phase, in which the CE is constant, while the specific heat fluctuates with $\bar{T}$ (this is not visible in Fig. \ref{CvFig}-(A), but can be easily verified -- we will discuss this point in detail in Section \ref{SecGSI}). Our conclusion is that CE is, in fact, directly correlated to the degrees of freedom of a system and how their dynamics affect the energy distribution. As the temperature increases and the system gains mass in the plasma phase, the total energy in the grand-canonical ensemble tends to stabilize, which can be observed in Fig. \ref{FBHEs}-(B), where $E_{BH}(\bar{T}, \omega l)$ is described by a flat surface in the high-temperature limit, for states rotating at speeds considerably lower than the speed of light. This type of stabilization is in complete harmony with the stability predicted by the CE, which does not necessarily correspond to the thermodynamic stability of a state described by its specific heat, as we will see more clearly in the next section.

\subsection{Crossovers}
\label{SecCross}
In the previous section, we have analyzed the CE for a system composed of a single state. In a more general case, we can consider an ensemble of particles rotating with different rotational velocities. This system can be holographically represented by a combination of charged BHs with different angular momentum. In the 3D-QCD phase diagram, it could describe the dynamics of the system at a given quark chemical potential, in particular, at $\bar{\mu}_0$, with the deconfinement temperature $T_c(\bar{\mu}_0, \omega l)$ being determined by the strongly interacting $N_p$ particles of the ensemble with different rotational velocities, according to the interval defined by Eq. \eqref{deltaw}. In this case, taking into account relativistic rotation effects (with $\omega l \geq 16\%$ of the speed of light), the system is given by a mixed phase, where hadronic and plasma states can coexist, see \cite{Junqueira:2025evt}. Applying the same method, with $\omega \in [0.3,\,0.4]$, the total energy density in the large $N_p$ limit, in Fourier space, reads         
\begin{equation}
\tilde{\rho}_{t}(k, \bar{T}) = \frac{1}{a_\omega}\left[\int_{0.3}^{0.4}d\omega\, \tilde{\rho}_{AdS}(k, \bar{T}, \omega l) + \int_{0.4}^{0.8}d\omega\, \tilde{\rho}_{BH}(k, \bar{T}, \omega l)\right]\;,
\end{equation}
so that, using Eqs. \eqref{tilderhoAdS} and \eqref{tilderhoBH}, it yields 
\begin{equation}\label{tilderhot}
    \!\!\tilde{\rho}_{t}(k, \bar{T}) \!=\! \frac{1}{a_\omega \pi^{3/2}}\!\left[\left( \frac{\sqrt{\pi}}{2}e^{-k^2/4}\!+\!i\mathcal{D}_+\!\left(\frac{k}{2}\right)\right)\!X_{AdS}( \bar{T})+\int_{0.4}^{0.8}\!\!d\omega \left[U(k,\bar{z}_h) X_{BH}(\bar{T},\omega l, \bar{z}_h)\!+\!iV(k,\bar{z}_h) X_{BH}(\bar{T},\omega l, \bar{z}_h)\right]\right]
\end{equation}
where we have defined
\begin{eqnarray}
X_{AdS}(\bar{T}) &=& \int_{0.3}^{0.4}d\omega\, E_{AdS}(\bar{T}, \omega l)\;,\\
X_{BH}(\bar{T},\omega l, \bar{z}_h) &=& \frac{E_{BH}(\bar{T}, \omega l)}{\text{Erf}(\bar{z}_h)}\;,\\
U(k, \bar{z}_h) &=& \int_0^{\bar{z}_h} dz\, e^{-\bar{z}^2} \cos(k \bar{z})\;,\label{U}\\
V(k, \bar{z}_h) &=& \int_0^{\bar{z}_h} dz\,e^{-\bar{z}^2} \sin(k \bar{z})\;.\label{V}
\end{eqnarray}

It is important to note that we cannot integrate $ X _ {BH} $ directly over angular momentum, since $U$ and $V$ also depend on $\omega l$. This dependence appears in the upper limit of the integration over the Poincaré coordinate, as the horizon position depends on the BH rotational velocity. As the integrals present in Eqs. \eqref{U} and \eqref{V} do not have an analytical solution, we must apply numerical methods to Eq. \eqref{tilderhot} to compute the double integration over $z$ and angular momentum. According to this equation, the absolute value of the energy density of the ensemble reads 
\begin{eqnarray}
     \vert \tilde{\rho}^{(a)}_{t}(k, \bar{T})\vert^2 &=& \left(\frac{\sqrt{\pi}}{2}e^{-k^2/4}X_{AdS}(\bar{T})+\int_{0.4}^{0.8}d\omega \, U(k,\bar{z}_h) X_{BH}(\bar{T},\omega l, \bar{z}_h) \right)^2 \nonumber\\ 
     && +\left(\mathcal{D}_+(\frac{k}{2})X_{AdS}( \bar{T}) + \int_{0.4}^{0.8}d\omega \,V(k,\bar{z}_h) X_{BH}(\bar{T},\omega l, \bar{z}_h)\right)^2\;,
\end{eqnarray}
where we have defined 
\begin{equation}
\tilde{\rho}^{(a)}_{t}(k,\bar{T}) = a_\omega \pi^{3/2} \tilde{\rho}_{t}(k,\bar{T})\;.
\end{equation}
This definition is useful since the modal fraction $f^{(a)}(k,\bar{T}) =f(k,\bar{T})$, since $\mathcal{N}^{(a)}(k,\bar{T}) = a_\omega \pi^{3/2} \mathcal{N}(k,\bar{T})$,
which shows that the ensemble CE does not depend on the parameter $a_\omega$.

\begin{figure}[!htb]
	\centering
    \includegraphics[scale=0.54]{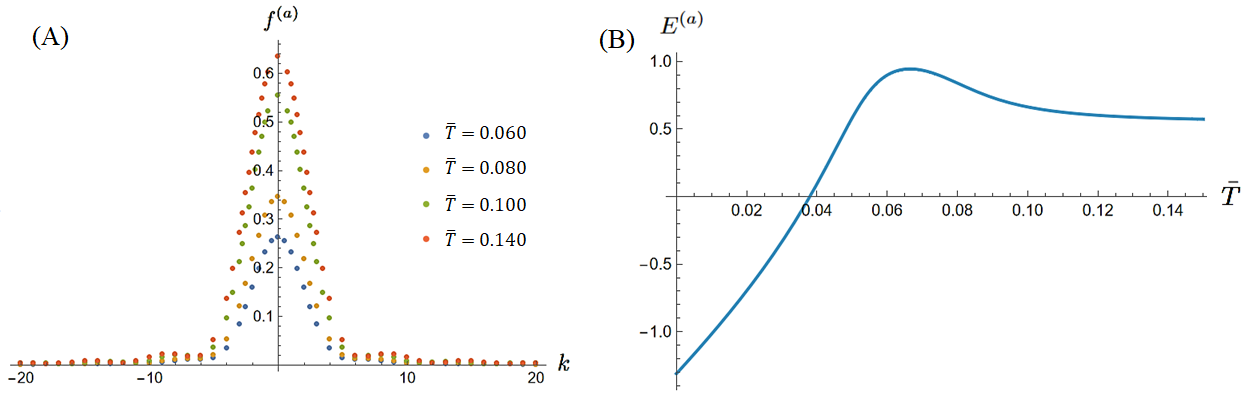}
	\caption{(A) Function $f^{(a)}(k)$ for the ensemble at different temperatures, and (B) total energy in the grand-canonical ensemble at  $\bar{\mu} = 0.396$, as a function of $\bar{T}$.}
    \label{fa}
\end{figure}

\begin{figure}[!htb]
	\centering
    \includegraphics[scale=0.54]{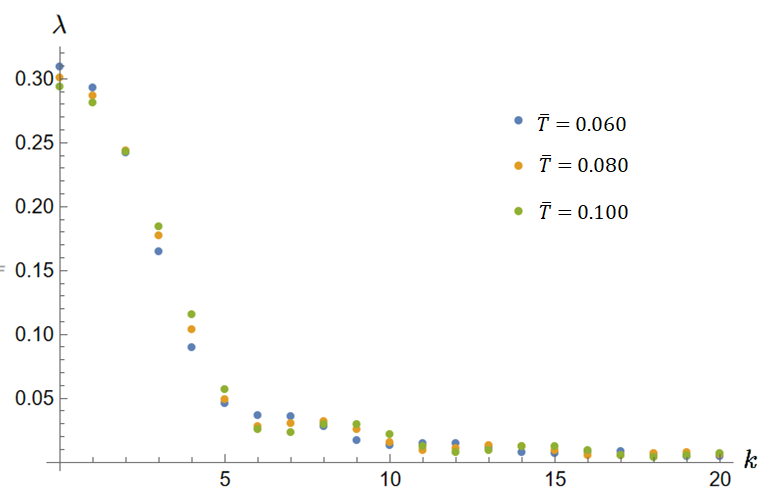}
	\caption{Function $\lambda(k)$ for the ensemble at different temperatures.}
    \label{figlambda}
\end{figure}

\begin{figure}[!htb]
	\centering
    \includegraphics[scale=0.5]{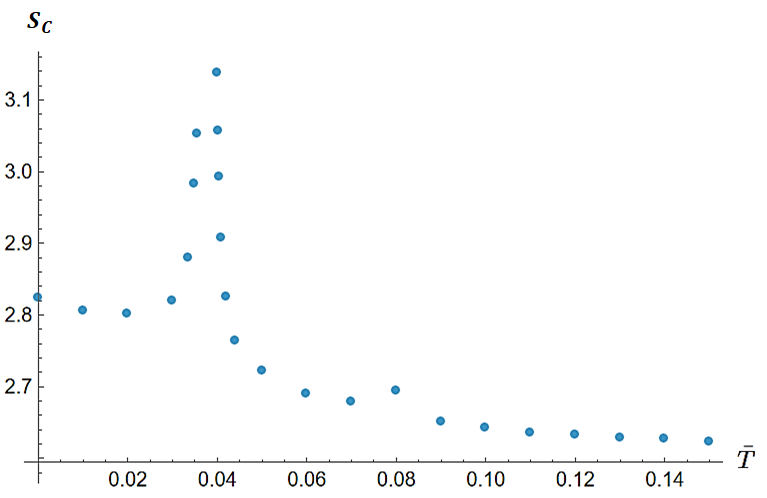}
	\caption{CE of the ensemble as a function of $\bar{T}$. }
    \label{CEens}
\end{figure}

From Fig. \ref{fa}-(A), see table \ref{table2}, one concludes that the ensemble modal fraction is finite and tends to zero as $k \rightarrow \pm \infty$. In Fig. \ref{fa}-(B), we have plotted the total energy of the ensemble, for $E^{(a)} = a_\omega E$, which shows that $E$ also does not diverge. These results show that the system obeys the \textit{first condition} for the modal fraction. The \textit{second condition} is also satisfied, as we can see in Fig. \ref{figlambda}, whre we have used table \ref{table3}. The CE for the ensemble of particles rotating with different angular momentum is plotted in Fig. \ref{CEens}, according to table \ref{table4}. As we can see, the CE is smooth along the entire temperature axis. In the holographic description of the QCD phase diagram obtained by the exact Andreev soft wall model for a strongly interacting rotating matter, this type of ensemble appears in the low-density and high-temperature regime, due to the effect of relativistic rotations ($\omega l \geq 0.16\%$ of the speed of light), see \cite{Junqueira:2025evt}. In this regime, the transition is characterized by \textit{crossovers}, which are consistently described by a CE without gaps.  

Comparing the CE curve with Fig. \ref{CvFig}-(B), one concludes that there is no direct correspondence between the CE and the thermodynamic stability of the system. The CE curve shows a peak at $\bar{T}$ slightly less than $0.04$, indicating instability of the system according to the Gleiser--Stamatopoulos criterion. In the same region, the specific heat is increasing, indicating that the system is thermodynamically stable, describing a progressive increase in its mass. It demonstrates that the informational and thermodynamic instabilities have distinct origins which cannot always be directly correlated.  

\section{Thermodynamic instability and Gleiser--Stamatopoulos interpretation}
\label{SecGSI}

\begin{figure}[!htb]
	\centering
    \includegraphics[scale=0.54]{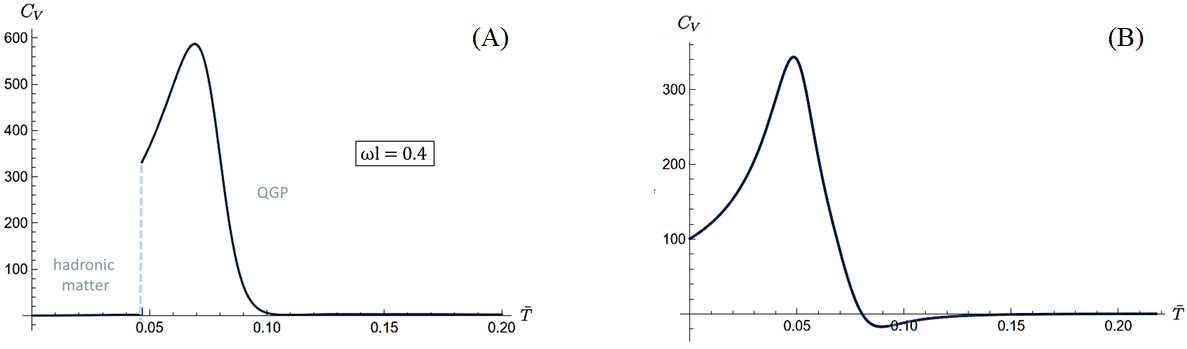}
	\caption{Specific heat as a function of $\bar{T}$: (A) for a single state rotating with $\omega l = 0 .4$; and (B) for an ensemble of particles with different angular momentum, for the rotational velocity varying in the interval $0.3 \leq \omega l \leq 0.8$. }
    \label{CvFig}
\end{figure}

If we compare the stability described by the CE with the thermodynamic one in the deconfined phase for the ensemble of particles rotating with different angular momentum, we observe a local maximum in Fig. \ref{CEens} for $\bar{T} \approx 0.08$, which corresponds to the temperature at which the specific heat becomes negative, see Fig. \ref{CvFig}-(B). This maximum only appears for the ensemble. The specific heat for a single state at a given rotational velocity is always positive, see \ref{CvFig}-(A), and its CE is always decreasing in the plasma phase, see Fig. \ref{SC1st}. In this case, the CE exhibits a good response to the thermodynamic instability of the system described by its negative specific heat. In any case, we must understand this result not as a necessary correlation, but as something that depends on how the loss/gain of mass affects the energy distribution. As soon as the system begins to lose energy (or mass), the total energy curve shows a steep decreasing slope until it stabilizes at the high-temperature limit. This stabilization is well described by the CE, since it tends towards a constant value as the temperature increases, see Fig. \ref{CEens}.

This interpretation can also be applied to the results obtained in Ref. \cite{Braga:2023qee}, where no direct correlation was found between the CE and the specific heat of the system. In this article, the CE increases sharply in the high-temperature limit. The same occurs with the specific heat. While the CE indicates matter instability in this temperature range, the specific heat points to thermodynamic stability. In fact, the mass gain (positive specific heat) yields a high slope of the total energy, which would correspond to instability predicted by the CE, since any variation in the vibration modes would induce the system to change its state. We should therefore not impose a direct correlation between the instability predicted by the CE and the thermodynamic instability of the system described by its specific heat. This imposition would completely fail in the present case, where we observe that the CE presents a peak in a region where the specific heat increases continuously. We must also point out that we are not assuming, in this model, thermal equilibrium between the rotating black hole and an external radiation background.

The physical interpretation of the CE peak can be established from the results obtained in Ref. \cite{Junqueira:2025evt}. In this work, we demonstrate that the critical point of the QCD phase diagram in the exact Andreev model occurs at $\bar{z}_{h c} = 0.832166$ and $\bar{\mu}_{CP} = 0.414$, up to a value of the plasma rotation velocity at which the relativistic effect no longer promotes mixing between the phases ($\omega l = 0.30 \pm 0,01$). Using these values, one gets the following critical point temperature:
\begin{eqnarray}
    \bar{T}_{CP} = \bar{T}(0.832166,0.414, 0.30) \approx 0.0383016\;.  
\end{eqnarray}
From a numerical analysis of the CE near the peak, we obtain
\begin{eqnarray}
    S_C(\bar{T} = 0.0375) &=& 4.4933\;,\nonumber\\
     S_C(\bar{T} = 0.0380) &=& 5.3755\;,\nonumber\\
     S_C(\bar{T} = 0.0385) &=& 5.0639\;,\nonumber\\
     S_C(\bar{T} = 0.0390) &=& 4.0309\;,
\end{eqnarray}
which shows that the CE peak occurs at $\bar{T} \approx 0.038$. One concludes, neglecting numerical uncertainties ($\delta S_C = \pm 0.0001$), that this peak corresponds to the critical point of the QCD phase diagram,
\begin{eqnarray}
    S_C\vert_{max} = S_C(\bar{T}_{CP})\;. 
\end{eqnarray}

The pronounced maximum of the CE near the critical
region is accompanied by a change in sign of the
total energy of the ensemble at $\bar{T}\approx 0.038$.
The proximity of these two features can be investigated
directly from the spectral structure of the ensemble
energy distribution. According to the Fourier
convention adopted here, the zero mode of the radial
energy distribution satisfies
\begin{equation}\label{ZeroModeEnergy}
    \tilde{\rho}_{\mathrm{ens}}(0)
    =
    \frac{1}{2\pi}
    \int_0^\infty
    \rho_{\mathrm{ens}}(\bar{z})\,d\bar{z}
    =
    \frac{E_{\mathrm{ens}}}{2\pi}\;.
\end{equation}
The corresponding modal fraction at zero radial
wavenumber consequently reads
\begin{equation}\label{ZeroModalFraction}
    f_{\mathrm{ens}}(0)
    =
    \frac{E_{\mathrm{ens}}^2}
    {4\pi^2\mathcal{N}_{\mathrm{ens}}}\;.
\end{equation}
Hence, provided that the normalization remains finite
and nonvanishing, the zero crossing of the total energy
necessarily entails the suppression of the zero mode
of the Fourier spectrum.

The spectral consequences of this cancellation can be
further elucidated by expanding the Fourier transform
around $k=0$,
\begin{equation}\label{LowKExpansion}
    \tilde{\rho}_{\mathrm{ens}}(k)
    =
    \frac{1}{2\pi}
    \left[
        E_{\mathrm{ens}}
        +ikM_1
        -\frac{k^2}{2}M_2
        +\mathcal{O}(k^3)
    \right]\;,
\end{equation}
where
\begin{equation}\label{RadialMoments}
    M_n =
    \int_0^\infty
    \bar{z}^{\,n}\rho_{\mathrm{ens}}(\bar{z})
    \,d\bar{z}
\end{equation}
denotes the $n^{\rm th}$  radial moment of the ensemble
energy distribution. At the zero crossing of the
total energy, the reality of the radial energy
distribution implies that the modal fraction
exhibits the low-$k$ behavior
\begin{equation}\label{InfraredModalSuppression}
    f_{\mathrm{ens}}(k)
    =
    \frac{M_1^2}
    {4\pi^2\mathcal{N}_{\mathrm{ens}}}
    k^2+\mathcal{O}(k^4)\;,
\end{equation}
provided that $M_1\neq0$. The cancellation of the
integrated energy therefore induces a quadratic
suppression of the modal fraction in the vicinity
of $k=0$, modifying the relative spectral weights
of the remaining Fourier components.

In addition, decomposing the ensemble energy distribution
into its hadronic and QGP contributions, the corresponding
Fourier power spectrum can be expressed as
\begin{align}\label{SpectralCrossTerms}
    |\tilde{\rho}_{\mathrm{ens}}(k)|^2
    &=
    |\tilde{\rho}_{\mathrm{had}}(k)|^2
    +
    |\tilde{\rho}_{\mathrm{QGP}}(k)|^2
    +
    2\operatorname{Re}
    \left[
        \tilde{\rho}_{\mathrm{had}}(k)
        \tilde{\rho}_{\mathrm{QGP}}^*(k)
    \right]\;.
\end{align}
The spectral cross term describes the constructive
and destructive interference between the Fourier
amplitudes of the confined and deconfined radial
energy distributions. Near the zero crossing of the
total energy, their integrated contributions compensate
each other, while their distinct radial profiles
generally prevent a complete cancellation of the
finite-$k$ Fourier components.

The resulting redistribution of spectral weights
may contribute to the increase of the CE in this
temperature interval. Although the suppression of
the zero mode does not, by itself, constitute a
sufficient condition for the emergence of a CE
maximum, the numerical proximity between these
two features suggests a connection between the
configurational reorganization of the ensemble
and the competition between its confined and
deconfined contributions.
The absence of a corresponding anomaly in the specific
heat indicates that the observed spectral reorganization
is not directly reflected in this thermodynamic response
function. Within the adopted radial prescription and
renormalization scheme, the proximity between the CE
maximum and the zero crossing of the ensemble energy,
together with the associated suppression of the low-$k$
modal fraction, provides a complementary
information-theoretic characterization of the
region separating first-order transitions from
crossover-like behavior. Whether this correspondence
persists under alternative radial prescriptions and
renormalization schemes remains an open question.

\section{Conclusions}
\label{SecConclusions}

We have investigated the configurational properties of
rotating strongly interacting matter at finite density
within Andreev's soft-wall AdS/QCD model. Holographic
renormalization allowed us to obtain the individual
thermodynamic energies of the confined and deconfined
phases, from which their radial energy distributions
and corresponding CEs were constructed. Within the adopted radial prescription, the CE remains
constant in the hadronic phase, independently of the
temperature, chemical potential, and angular velocity.
This property follows from the invariance of the modal
fraction under overall energy rescalings. The discontinuity
of the CE across first-order confinement/deconfinement
transitions contrasts with its continuous behavior for
ensembles with distributed angular velocities, providing
an informational characterization of crossover-like
behavior in the holographic QCD phase diagram.

Ref. \cite{Junqueira:2025evt} demonstrated the emergence of a critical point in Andreev's exact model when relativistic rotational effects are taken into account. In the low-density regime, the system is characterized by a mixed phase, with the transition occurring smoothly. In the high-density regime, the curves are dominated by the critical temperature of nonrotating matter. The former regime corresponds to an ensemble of particles rotating with distinct angular momenta, characterized by a continuous CE without gaps, as shown in Fig. \ref{CEens}. The latter corresponds to a first-order phase transition at fixed angular momentum ($\omega l \approx 0$), consistently emulating the HP transition depicted in Fig. \ref{SC1st}, where a gap emerges. These results indicate that the nature of the phase transition in the QCD phase diagram is correlated with how information is redistributed within the system, as encoded in the corresponding CE profiles, with or without gaps.

Within the confined phase, the CE remains constant, providing a characteristic signature of color confinement. The same behavior was reported in the hard-wall model in Ref. \cite{Braga:2020opg}, indicating that this feature is insensitive to the particular implementation of the IR scale in the models considered. In addition, as follows from Eq. \eqref{FAdS}, this behavior is independent of the angular momentum of the particles. Consequently, Fig. \ref{FEs}-(A) holds for any admissible value of $\omega l$. Ref. \cite{BallonBayona:2007vp} argued that the confinement/deconfinement transition can be identified by a discontinuity in the thermodynamic entropy, from $S_{\mathrm{AdS}} = 0$ to $S_{\mathrm{BH}} \sim N_c^2 T^3$. However, in the rotating extension of the exact model, the confined phase is no longer characterized by vanishing thermodynamic entropy.  Indeed, as shown in Fig. 9 of Ref. \cite{Junqueira:2025evt}, the thermodynamic entropy varies within the confined phase due to the additional degrees of freedom associated with the angular momentum of the particles. In this respect, the constancy of the CE provides a more robust criterion for identifying color confinement within the holographic models considered, exhibiting the same qualitative behavior for both rotating and nonrotating matter.

The comparison with the specific heat reveals that
configurational and thermodynamic stability need not
exhibit a direct correspondence. Although a local maximum
of the ensemble CE occurs near the onset of negative
specific heat, a more pronounced maximum emerges at
$\bar{T}\approx 0.038$, without a corresponding anomaly
in the specific heat. Its proximity to the critical
region of the holographic QCD phase diagram suggests
that the CE is sensitive to spectral rearrangements
associated with changes in the nature of the
confinement/deconfinement transition. 
Remarkably, this maximum occurs in the vicinity of the
zero crossing of the ensemble energy. We have analytically
established that the cancellation of the integrated
energy suppresses the zero Fourier mode and, for a
nonvanishing first radial moment, induces a quadratic
suppression of the modal fraction at small radial
wavenumbers. The spectral interference between the
hadronic and plasma contributions provides a possible
mechanism underlying the observed enhancement of the CE.

These results establish an information-theoretic perspective
on the phase structure of rotating strongly interacting
matter, emphasizing the use of the CE to distinguish first-order confinement/deconfinement
transitions from crossover-like behavior and to characterize
the critical region of the holographic QCD phase diagram.
The interplay between thermodynamic observables and
configurational information provides a complementary
setup for investigating the spectral reorganization
of strongly interacting matter across the hadronic and
QGP phases. Further developments along
these lines may deepen our understanding of QCD phase
transitions from the point of view of information theory,
particularly in regimes where conventional thermodynamic
response functions exhibit no pronounced signatures.
The robustness of these findings under alternative radial
prescriptions and holographic renormalization procedures,
as well as their connection with the dynamical stability
of the QGP, deserves further investigation.

\subsection*{Acknowledgements}
OCJ thanks the S\~ao Paulo Research Foundation (FAPESP) 
(Grants No. 2021/01089-1 and No. 2024/14390-0). The work of RdR is supported by FAPESP  (Grants No. 2021/01089-1 and No. 2026/14943-4) and the National Council for Scientific and Technological Development - CNPq  (Grants No. 303742/2023-2 and No. 401567/2023-0).

\appendix

\section{Auxiliary tables with numerical results} \label{app1}
This appendix collects supplementary numerical results underlying the 
CE analysis presented in the main text. 
Table~\ref{table1} lists the black hole modal fraction normalizations 
$\mathcal{N}_{BH}$ and the corresponding CEs  
$S_C$ as functions of the dimensionless temperature $\bar{T}$, 
at fixed rotational parameter $\omega l=0.4$.

\begin{table}[H]
\centering
\begin{tabular}[c]{|c|!{\vrule width 1pt}c|c|}
\hline 
\text{}   $\,\,\,\bar{T}\,\,\, $ & $\,\,\,\mathcal{N}_{BH}\,\,\, $& $\,\,\,S_C\,\,\, $\\
 \hline\hline
 $\,\,\,0.047\,\,\, $ &$\,\,\,23.0996\,\,\,$ &$\,\,\,1.77789\,\,\,$ \\
\hline
  $\,\,\,0.055\,\,\, $ &$\,\,\,25.5041\,\,\,$&$\,\,\,1.74935\,\,\,$   \\
\hline
  $\,\,\,0.065\,\,\, $ &$\,\,\,29.0924\,\,\,$  &$\,\,\,1.71374\,\,\,$ \\
\hline
 $\,\,\,0.075\,\,\, $ &$\,\,\,31.3794\,\,\,$&$\,\,\,1.67332\,\,\,$  \\
\hline
  $\,\,\,0.085\,\,\, $ &$\,\,\,27.3543\,\,\,$ &$\,\,\,1.62925\,\,\,$ \\
\hline
 $\,\,\,0.095\,\,\, $ &$\,\,\,21.0352\,\,\,$&$\,\,\,1.59847\,\,\,$  \\
\hline
 $\,\,\,0.105\,\,\, $ &$\,\,\,16.9786\,\,\,$ &$\,\,\,1.57506\,\,\,$ \\
\hline
 $\,\,\,0.115\,\,\, $ &$\,\,\,14.6252\,\,\,$&$\,\,\,1.55893\,\,\,$  \\
\hline
 $\,\,\,0.125\,\,\, $ &$\,\,\,13.1598\,\,\,$&$\,\,\, 1.54732\,\,\,$  \\
\hline
 $\,\,\,0.135\,\,\, $ &$\,\,\,12.1692\,\,\,$&$\,\,\,1.53681\,\,\,$  \\
\hline
 $\,\,\,0.145\,\,\, $ &$\,\,\,11.4541\,\,\,$&$\,\,\, 1.52723\,\,\,$  \\
\hline
 $\,\,\,0.155\,\,\, $ &$\,\,\,10.9129\,\,\,$&$\,\,\,1.52063\,\,\,$  \\
\hline
 $\,\,\,0.165\,\,\, $ &$\,\,\,10.4856\,\,\,$&$\,\,\,1.51417\,\,\,$  \\
\hline
  $\,\,\,0.175\,\,\, $ &$\,\,\,10.1388\,\,\,$ &$\,\,\,1.50919\,\,\,$ \\
\hline
 $\,\,\,0.185\,\, $ &$\,\,\,9.84916\,\,\,$&$\,\,\,1.50370\,\,\,$   \\
\hline
  $\,\,\,0.195\,\,\, $ &$\,\,\,9.60434\,\,\,$&$\,\,\,1.50062\,\,\,$   \\
\hline
 $\,\,\,0.205\,\,\, $ &$\,\,\,9.39192\,\,\,$&$\,\,\,1.49653\,\,\,$  \\
\hline
 \end{tabular}   
\caption{Black hole modal fraction normalizations and CEs, for a BH rotating with $\omega l = 0.4$, used to plot Fig. \ref{SC1st}. }
\label{table1}
\end{table} 

Table~\ref{table2} provides numerical values of the modal fraction 
$f^{(a)}(k)$ at selected temperatures, resolving the redistribution 
of the spectral weights across momentum modes. These data complement 
the graphical results and facilitate their numerical reproducibility.
\begin{table}[H]
\centering
\begin{tabular}[c]{|c|!{\vrule width 1pt}c|c|c|c|}
\hline 
\text{}   $\,\,\,k\,\,\, $ & $\,\,\,f^{(a)}_{\bar{T}_1}\,\,\, $& $\,\,\,f^{(a)}_{\bar{T}_2}\,\,\, $& $\,\,\,f^{(a)}_{\bar{T}_3}\,\,\, $& $\,\,\,f^{(a)}_{\bar{T}_4}\,\,\, $\\
 \hline\hline
 $\,\,\,0.00\,\,\, $ &$\,\,\,0.634564\,\,\,$ &$\,\,\,0.554064\,\,\,$&$\,\,\,0.345543\,\,\,$ &$\,\,\,0.263004\,\,\,$ \\
\hline
  $\,\,\,0.50\,\,\, $ &$\,\,\,0.619867\,\,\,$&$\,\,\,0.539621\,\,\,$&$\,\,\,0.335669\,\,\,$ &$\,\,\,0.254867\,\,\,$   \\
\hline
  $\,\,\,0.75\,\,\, $ &$\,\,\,0.601997\,\,\,$&$\,\,\,0.522106\,\,\,$&$\,\,\,0.323728\,\,\,$ &$\,\,\,0.245050\,\,\,$   \\
\hline
  $\,\,\,1.00\,\,\, $ &$\,\,\,0.577877\,\,\,$&$\,\,\,0.498549\,\,\,$  &$\,\,\,0.307729\,\,\,$ &$\,\,\,0.231939\,\,\,$ \\
\hline
  $\,\,\,1.25\,\,\, $ &$\,\,\,0.548321\,\,\,$  &$\,\,\,0.469827\,\,\,$&$\,\,\,0.288321\,\,\,$ &$\,\,\,0.216106\,\,\,$ \\
\hline
 $\,\,\,1.50\,\,\, $ &$\,\,\,0.514277\,\,\,$&$\,\,\,0.436953\,\,\,$ &$\,\,\,0.266251\,\,\,$ &$\,\,\,0.198206\,\,\,$ \\
\hline
  $\,\,\,1.75\,\,\, $ &$\,\,\,0.476767\,\,\,$ &$\,\,\,0.401018\,\,\,$&$\,\,\,0.242316\,\,\,$ &$\,\,\,0.178932\,\,\,$ \\
\hline
 $\,\,\,2.00\,\,\, $ &$\,\,\,0.436831\,\,\,$&$\,\,\,0.363127\,\,\,$ &$\,\,\,0.217313\,\,\,$ &$\,\,\,0.158976\,\,\,$ \\
\hline
 $\,\,\,2.25\,\,\, $ &$\,\,\,0.395488\,\,\,$ &$\,\,\,0.324349\,\,\,$&$\,\,\,0.192005\,\,\,$ &$\,\,\,0.138988\,\,\,$ \\
\hline
 $\,\,\,2.50\,\,\, $ &$\,\,\,0.353698\,\,\,$&$\,\,\,0.285679\,\,\,$ &$\,\,\,0.167084\,\,\,$ &$\,\,\,0.119550\,\,\,$ \\
\hline
 $\,\,\,2.75\,\,\, $ &$\,\,\,0.312340\,\,\,$&$\,\,\, 0.248005\,\,\,$&$\,\,\,0.143153\,\,\,$ &$\,\,\,0.101153\,\,\,$  \\
\hline
 $\,\,\,3.00\,\,\, $ &$\,\,\,0.272191\,\,\,$&$\,\,\,0.212091\,\,\,$ &$\,\,\,0.120710\,\,\,$ &$\,\,\,0.0841903\,\,\,$ \\
\hline
 $\,\,\,3.50\,\,\, $ &$\,\,\,0.198102\,\,\,$&$\,\,\, 0.147923\,\,\,$ &$\,\,\,0.0817381\,\,\,$ &$\,\,\,0.0556045\,\,\,$ \\
\hline
 $\,\,\,4.00\,\,\, $ &$\,\,\,0.135458\,\,\,$&$\,\,\,0.0965403\,\,\,$ &$\,\,\,0.0519725\,\,\,$ &$\,\,\,0.0348598\,\,\,$ \\
\hline
 $\,\,\,5.00\,\,\, $ &$\,\,\,0.0515946\,\,\,$&$\,\,\,0.0350787\,\,\,$ &$\,\,\,0.0195903\,\,\,$ &$\,\,\,0.0144616\,\,\,$ \\
\hline
  $\,\,\,6.00\,\,\, $ &$\,\,\,0.0186617\,\,\,$ &$\,\,\,0.0173130\,\,\,$ &$\,\,\,0.0123864\,\,\,$ &$\,\,\,0.0108316\,\,\,$\\
\hline
 $\,\,\,7.00\,\, $ &$\,\,\,0.0166065\,\,\,$&$\,\,\,0.0190785\,\,\,$&$\,\,\,0.0132907\,\,\,$ &$\,\,\,0.0104064\,\,\,$   \\
\hline
  $\,\,\,8.00\,\,\, $ &$\,\,\,0.0222779\,\,\,$&$\,\,\,0.0204751\,\,\,$&$\,\,\,0.0117962\,\,\,$ &$\,\,\,0.00762633\,\,\,$   \\
\hline
 $\,\,\,9.00\,\,\, $ &$\,\,\,0.0222779\,\,\,$&$\,\,\,0.0153492\,\,\,$ &$\,\,\,0.00731228\,\,\,$ &$\,\,\,0.00424236\,\,\,$ \\
\hline
  $\,\,\,10.0\,\,\, $ &$\,\,\,0.0220712\,\,\,$ &$\,\,\,0.00813348\,\,\,$&$\,\,\,0.00382544\,\,\,$ &$\,\,\,0.00292916\,\,\,$ \\
\hline
 $\,\,\,11.0\,\, $ &$\,\,\,0.00732448\,\,\,$&$\,\,\,0.00449866\,\,\,$&$\,\,\,0.00335102\,\,\,$ &$\,\,\,0.00334816\,\,\,$   \\
\hline
  $\,\,\,12.0\,\,\, $ &$\,\,\,0.00399481\,\,\,$&$\,\,\,0.00518445\,\,\,$ &$\,\,\,0.00429248\,\,\,$ &$\,\,\,0.00347618\,\,\,$  \\
\hline
 $\,\,\,13.0\,\,\, $ &$\,\,\,0.00517498\,\,\,$&$\,\,\,0.00682276\,\,\,$ &$\,\,\,0.00428152\,\,\,$ &$\,\,\,0.00254691\,\,\,$ \\
\hline
  $\,\,\,14.0\,\,\, $ &$\,\,\,0.00738123\,\,\,$ &$\,\,\,0.00644666\,\,\,$&$\,\,\,0.00295072\,\,\,$ &$\,\,\,0.00153565\,\,\,$ \\
\hline
 $\,\,\,15.0\,\, $ &$\,\,\,0.00750647\,\,\,$&$\,\,\,0.00422719\,\,\,$ &$\,\,\,0.00168976\,\,\,$ &$\,\,\,0.00136349\,\,\,$  \\
\hline
  $\,\,\,16.0\,\,\, $ &$\,\,\,0.00531518\,\,\,$&$\,\,\,0.00233550\,\,\,$ &$\,\,\,0.00153715\,\,\,$ &$\,\,\,0.00169992\,\,\,$  \\
\hline
 $\,\,\,17.0\,\,\, $ &$\,\,\,0.00281234\,\,\,$&$\,\,\,0.00214657\,\,\,$ &$\,\,\,0.00205457\,\,\,$ &$\,\,\,0.00169386\,\,\,$ \\
\hline
$\,\,\,18.0\,\, $ &$\,\,\,0.00186690\,\,\,$&$\,\,\,0.00300755\,\,\,$&$\,\,\,0.00218480\,\,\,$ &$\,\,\,0.00119811\,\,\,$   \\
\hline
  $\,\,\,19.0\,\,\, $ &$\,\,\,0.00257348\,\,\,$&$\,\,\,0.00339830\,\,\,$ &$\,\,\,0.00161915\,\,\,$ &$\,\,\,0.000788531\,\,\,$  \\
\hline
 $\,\,\,20.0\,\,\, $ &$\,\,\,0.00358101\,\,\,$&$\,\,\,0.00270554\,\,\,$&$\,\,\,0.000980147\,\,\,$ &$\,\,\,0.000832519\,\,\,$  \\
\hline
 \end{tabular}   
\caption{Function $f^{(a)}(k)$ at different temperatures ($\bar{T}_1 = 0.060$, $\bar{T}_2 = 0.080$, $\bar{T}_3 = 0.100$ e $T_4 = 0.140$), used to plot Fig. \ref{fa}-(A). }
\label{table2}
\end{table}

\begin{table}[H]
\centering
\begin{tabular}[c]{|c|!{\vrule width 1pt}c|c|c|}
\hline 
\text{}   $\,\,\,k\,\,\, $ & $\,\,\,\lambda_{\bar{T}_1}\,\,\, $& $\,\,\,\lambda_{\bar{T}_2}\,\,\, $& $\,\,\,\lambda_{\bar{T}_3}\,\,\, $\\ 
 \hline\hline
 $\,\,\,0.00\,\,\, $ &$\,\,\,0.293114\,\,\,$ &$\,\,\,0.300110\,\,\,$&$\,\,\,0.304935\,\,\,$  \\
\hline
  $\,\,\,1.00\,\,\, $ &$\,\,\,0.280554\,\,\,$&$\,\,\,0.286052\,\,\,$  &$\,\,\,0.289615\,\,\,$ \\
\hline
 $\,\,\,2.00\,\,\, $ &$\,\,\,0.242874\,\,\,$&$\,\,\,0.243364\,\,\,$ &$\,\,\,0.242783\,\,\,$  \\
\hline
 $\,\,\,3.00\,\,\, $ &$\,\,\,0.183777\,\,\,$&$\,\,\,0.176836\,\,\,$ &$\,\,\,0.170778\,\,\,$ \\
\hline
 $\,\,\,4.00\,\,\, $ &$\,\,\,0.115275\,\,\,$&$\,\,\,0.103608\,\,\,$ &$\,\,\,0.0956955\,\,\,$ \\
\hline
 $\,\,\,5.00\,\,\, $ &$\,\,\,0.0564547\,\,\,$&$\,\,\,0.0484501\,\,\,$ &$\,\,\,0.045745\,\,\,$ \\
\hline
  $\,\,\,6.00\,\,\, $ &$\,\,\,0.0252012\,\,\,$ &$\,\,\,0.0276314\,\,\,$ &$\,\,\,0.0317972\,\,\,$ \\
\hline
 $\,\,\,7.00\,\, $ &$\,\,\,0.022914\,\,\,$&$\,\,\,0.0298857\,\,\,$&$\,\,\,0.0336446\,\,\,$   \\
\hline
  $\,\,\,8.00\,\,\, $ &$\,\,\,0.0290904\,\,\,$&$\,\,\,0.0316333\,\,\,$&$\,\,\,0.0305736\,\,\,$    \\
\hline
 $\,\,\,9.00\,\,\, $ &$\,\,\,0.0288723\,\,\,$&$\,\,\,0.0250595\,\,\,$ &$\,\,\,0.0207219\,\,\,$ \\
\hline
  $\,\,\,10.0\,\,\, $ &$\,\,\,0.0211491\,\,\,$ &$\,\,\,0.0148503\,\,\,$&$\,\,\,0.0120951\,\,\,$  \\
\hline
 $\,\,\,11.0\,\, $ &$\,\,\,0.0116171\,\,\,$&$\,\,\,0.00902421\,\,\,$&$\,\,\,0.0108197\,\,\,$   \\
\hline
  $\,\,\,12.0\,\,\, $ &$\,\,\,0.0069462\,\,\,$&$\,\,\,0.0101761\,\,\,$ &$\,\,\,0.0133215\,\,\,$   \\
\hline
 $\,\,\,13.0\,\,\, $ &$\,\,\,0.00866078\,\,\,$&$\,\,\,0.0128219\,\,\,$ &$\,\,\,0.0132930\,\,\,$  \\
\hline
  $\,\,\,14.0\,\,\, $ &$\,\,\,0.0116928\,\,\,$ &$\,\,\,0.0122263\,\,\,$&$\,\,\,0.00971716\,\,\,$  \\
\hline
 $\,\,\,15.0\,\, $ &$\,\,\,0.0118593\,\,\,$&$\,\,\,0.00855969\,\,\,$ &$\,\,\,0.00604139\,\,\,$  \\
\hline
  $\,\,\,16.0\,\,\, $ &$\,\,\,0.00885962\,\,\,$&$\,\,\,0.00515073\,\,\,$ &$\,\,\,0.00556940\,\,\,$   \\
\hline
 $\,\,\,17.0\,\,\, $ &$\,\,\,0.00513881\,\,\,$&$\,\,\,0.00478913\,\,\,$ &$\,\,\,0.00714241\,\,\,$ \\
\hline
$\,\,\,18.0\,\, $ &$\,\,\,0.00360400\,\,\,$&$\,\,\,0.00640147\,\,\,$&$\,\,\,0.00752720\,\,\,$   \\
\hline
  $\,\,\,19.0\,\,\, $ &$\,\,\,0.00475990\,\,\,$&$\,\,\,0.00710690\,\,\,$ &$\,\,\,0.00582392\,\,\,$  \\
\hline
 $\,\,\,20.0\,\,\, $ &$\,\,\,0.00632534\,\,\,$&$\,\,\,0.00584575\,\,\,$&$\,\,\,0.00377449\,\,\,$  \\
\hline
 \end{tabular}   
\caption{Function $\lambda(k)$ for the ensemble at different temperatures ($\bar{T}_1$, $\bar{T}_2$ and $\bar{T}_3$, as defined in table \ref{table2}), used to plot Fig. \ref{figlambda}. }
\label{table3}
\end{table} 

\begin{table}[H]
\centering
\begin{tabular}[c]{|c|!{\vrule width 1pt}c|c|}
\hline 
\text{}   $\,\,\,\bar{T}\,\,\, $ & $\,\,\,\mathcal{N}^{(a)}\,\,\, $& $\,\,\,S_C\,\,\, $\\
 \hline\hline
 $\,\,\,0.00000\,\,\, $ &$\,\,\,10.1114\,\,\,$ &$\,\,\,2.82345\,\,\,$ \\
\hline
  $\,\,\,0.01000\,\,\, $ &$\,\,\,6.01922\,\,\,$&$\,\,\,2.80511\,\,\,$   \\
\hline
  $\,\,\,0.02000\,\,\, $ &$\,\,\,2.79481\,\,\,$  &$\,\,\,2.80157\,\,\,$ \\
\hline
 $\,\,\,0.03000\,\,\, $ &$\,\,\,0.652193\,\,\,$&$\,\,\,2.81899\,\,\,$  \\
\hline
  $\,\,\,0.03350\,\,\, $ &$\,\,\,0.228212\,\,\,$ &$\,\,\,2.87994\,\,\,$ \\
\hline
 $\,\,\,0.03500\,\,\, $ &$\,\,\,0.110273\,\,\,$&$\,\,\,2.98242\,\,\,$  \\
\hline
 $\,\,\,0.03550\,\,\, $ &$\,\,\,0.0802381\,\,\,$ &$\,\,\,3.05240\,\,\,$ \\
\hline
 $\,\,\,0.04000\,\,\, $ &$\,\,\,0.0424235\,\,\,$&$\,\,\,3.13689\,\,\,$  \\
\hline
 $\,\,\,0.04025\,\,\, $ &$\,\,\,0.0536861\,\,\,$&$\,\,\, 3.05638\,\,\,$  \\
\hline
 $\,\,\,0.04050\,\,\, $ &$\,\,\,0.0664365\,\,\,$&$\,\,\,2.99207\,\,\,$  \\
\hline
 $\,\,\,0.04100\,\,\, $ &$\,\,\,0.0965057\,\,\,$&$\,\,\, 2.90801\,\,\,$  \\
\hline
 $\,\,\,0.04200\,\,\, $ &$\,\,\,0.175316\,\,\,$&$\,\,\,2.82482\,\,\,$  \\
\hline
 $\,\,\,0.04400\,\,\, $ &$\,\,\,0.409956\,\,\,$&$\,\,\,2.76399\,\,\,$  \\
\hline
  $\,\,\,0.05000\,\,\, $ &$\,\,\,1.69038\,\,\,$ &$\,\,\,2.72259\,\,\,$ \\
\hline
 $\,\,\,0.06000\,\, $ &$\,\,\,3.96897\,\,\,$&$\,\,\,2.69008\,\,\,$   \\
\hline
  $\,\,\,0.07000\,\,\, $ &$\,\,\,4.18462\,\,\,$&$\,\,\,2.67821\,\,\,$   \\
\hline
 $\,\,\,0.08000\,\,\, $ &$\,\,\,3.28718\,\,\,$&$\,\,\,2.69445\,\,\,$  \\
\hline
 $\,\,\,0.09000\,\,\, $ &$\,\,\,2.44579\,\,\,$&$\,\,\,2.65026\,\,\,$  \\
\hline
  $\,\,\,0.10000\,\,\, $ &$\,\,\,1.97582\,\,\,$ &$\,\,\,2.64268\,\,\,$ \\
\hline
 $\,\,\,0.11000\,\, $ &$\,\,\,1.73060\,\,\,$&$\,\,\,2.63586\,\,\,$   \\
\hline
  $\,\,\,0.12000\,\,\, $ &$\,\,\,1.59289\,\,\,$&$\,\,\,2.63201\,\,\,$   \\
\hline
 $\,\,\,0.13000\,\,\, $ &$\,\,\,1.51057\,\,\,$&$\,\,\,2.62815\,\,\,$  \\
\hline
  $\,\,\,0.14000\,\,\, $ &$\,\,\,1.45700\,\,\,$&$\,\,\,2.62668\,\,\,$   \\
\hline
 $\,\,\,0.15000\,\,\, $ &$\,\,\,1.42076\,\,\,$&$\,\,\,2.62308\,\,\,$  \\
\hline
 \end{tabular}   
\caption{Modal fraction normalizations and CEs for the ensemble, used to plot Fig. \ref{CEens}. }
\label{table4}
\end{table} 
\bibliographystyle{utphys2}
\bibliography{library}

\end{document}